\documentclass[reprint,prx]{revtex4-1}

\usepackage{amsmath} % AMS Math Package
\usepackage{amsthm} % Theorem Formatting
\usepackage{amssymb}	% Math symbols such as \mathbb
\usepackage[pdftex]{graphicx}
\usepackage{epstopdf}
\usepackage{enumerate}
\usepackage{wrapfig}
\usepackage[dvips,letterpaper,margin=1in,bottom=1in]{geometry}
\usepackage{color}
\usepackage{tensind}
\tensordelimiter{`}
\whenindex{a}{\alpha}{} \whenindex{b}{\beta}{} \whenindex{g}{\gamma}{} \whenindex{d}{\delta}{} \whenindex{e}{\varepsilon}{} \whenindex{m}{\mu}{} \whenindex{n}{\nu}{} \whenindex{l}{\lambda}{} \whenindex{k}{\kappa}{} \whenindex{r}{\rho}{} \whenindex{s}{\sigma}{} \whenindex{t}{\tau}{}
\usepackage{todonotes}
\usepackage{subcaption}
\usepackage{setspace}
\usepackage{comment}
\usepackage{cancel}
\usepackage[]{units}

\usepackage{hyperref}
\usepackage[all]{hypcap}

\makeatletter % Need for anything that contains an @ command 

\newcommand{\pd}[2]{\frac{\partial #1}{\partial #2}} 
\let\baraccent=\= % rename builtin command \= to \baraccent
\renewcommand{\=}[1]{\stackrel{#1}{=}} % for putting numbers above =
\renewcommand{\(}{\left (}
\renewcommand{\)}{\right  )}
\renewcommand{\[}{\left [}
\renewcommand{\]}{\right ]}
\newcommand{\<}{\left <}
\renewcommand{\>}{\right >}

\theoremstyle{definition}

\theoremstyle{remark}

\newtheorem{?}{\textbf{Question}}

\newcommand{\e}[1]{\mbox{e}^{#1}} %exponential
\renewcommand{\exp}[1]{\mbox{exp}\[#1\]} %exponential
\newcommand{\D}{\mathcal{D}}
\newcommand{\Ord}[1]{O\[#1\]}  %order of (big O notation)
\newcommand{\Op}{\mathcal{O}}  %operator
\newcommand{\at}[1]{\,\big |_{#1}} %evaluate at value
\begin{document}
\title{Least Rattling Feedback from Strong Time-scale Separation}

\author{Pavel Chvykov}
\email{pchvykov@mit.edu}
\author{Jeremy England}
\affiliation{Physics of Living Systems, Massachusetts Institute of Technology} 
%		\thanks{pchvykov@mit.edu}

\begin{abstract}
In most interacting many-body systems associated with some ``emergent phenomena,'' we can identify sub-groups of degrees of freedom that relax on dramatically different time-scales. Time-scale separation of this kind is particularly helpful in nonequilibrium systems where only the fast variables are subjected to external driving; in such a case, it may be shown through elimination of fast variables that the slow coordinates effectively experience a thermal bath of spatially-varying temperature. In this work, we investigate how such a temperature landscape arises according to how the slow variables affect the character of the driven quasi-steady-state reached by the fast variables.  Brownian motion in the presence of spatial temperature gradients is known to lead to the accumulation of probability density in low temperature regions.  Here, we focus on the implications of attraction to low effective temperature for the long-term evolution of slow variables.  After quantitatively deriving the temperature landscape for a general class of overdamped systems using a path integral technique, we then illustrate in a simple dynamical system how the attraction to low effective temperature has a fine-tuning effect on the slow variable, selecting configurations that bring about exceptionally low force fluctuation in the fast-variable steady-state.  We furthermore demonstrate that a particularly strong effect of this kind can take place when the slow variable is tuned to bring about orderly, integrable motion in the fast dynamics that avoids thermalizing energy absorbed from the drive. We thus point to a potentially general feedback mechanism in multi-time-scale active systems, that leads to the exploration of slow variable space, as if in search of fine-tuning for a ``least rattling" response in the fast coordinates.

\end{abstract}

\maketitle

%\tableofcontents

%-appendix
%-abstract
%-SI
%-referees
%-go through references
\section{Introduction}

A broad range of many-body nonequilibrium systems have in common that different degrees of freedom within them undergo motion on two, well-separated time-scales, and that the faster degrees of freedom are the only ones directly subject to external driving.  Such separation can occur if a faster set of active particles act as a bath for a heavier, more slowly relaxing set of larger, extended degrees of freedom, such as in the example of a polymer immersed in a mixture of self-propelling particles \cite{nikola2016active_polymer}. Alternatively, in many systems one can usefully identify coarse-grained variables describing global features of the many-body dynamics, which may relax more slowly than the coordinates of individual particles. Such order parameters might then be thought of as a set of slowly-varying constraints on the driven fast dynamics, as for example in \cite{sasa2015kuramoto_thermo}.

In all such cases, it is possible in principle for the particular configuration of a set of slow variables to have a significant influence on the specific nonequilibrium steady-state reached by the fast variables.  Thus, in general, a feedback loop can arise in which the slow variables first establish the features of the fast steady-state, and then the statistics of this steady-state in turn determine the stochastic dynamics of the resulting local motion in slow variable space.  The goal of this paper is to characterize the dynamical attractors of slow variable evolution in terms of the particular, special properties of the fast steady-states to which they give rise.

Nonequilibrium systems with time-scale separation have been extensively studied over the last several decades. The most common context where they have come up is in formalizing the concept of a ``thermal bath'' -- explicitly modelling the fast bath degrees of freedom as a Hamiltonian system, and studying their effects on the slow variables. In this way, one can in some cases recover the effective friction tensor \cite{Berry1993chaotic_bath}, and the corresponding noise term, related by fluctuation-dissipation theorem \cite{jarzynski1995chotic_thermalization}. There is also extensive literature studying the conditions and effects of deviations from this basic result, which are generally termed ``anomalous diffusion'' -- see e.g., \cite{lutz2001fractional_Lang}. Within this context, the ``slow'' degrees of freedom lack their own dynamics, and are considered only as probes of the fast bath. More recent studies have considered the minimal dissipation required from an external agent to slowly move such probes. A geometric interpretation of this bound was presented in \cite{zulkowski2012geometry}, and extended to nonequilibrium baths in \cite{zulkowski2013neq_geom}, as well as to reversible external protocols in \cite{machta2015dissipation}. Systems where slow variables have their own dynamics under a conservative coupling to the fast bath have received relatively little attention, excepting notable recent work for a simple harmonic oscillator probe in \cite{polkovnikov2016neg_mass}, and a more general exploration in \cite{Maes2015neq_fluid,Maes2015Lin_t_seper}, where some formal results relating dissipation and forces on the slow variables were derived. 

Most of this previous work has relied on the projection operator technique to adiabatically eliminate fast variables and obtain the reduced Fokker-Planck equations for the slow variables, as in Ch6.4 of \cite{gardiner1985handbook}, or see \cite{bo2016time-sc_sep_functionals} for a recent review. The straightforward implication of this approach is that at long times, probability density in slow variable space is expected to accumulate in locations where inward mean drift is strong, and where local diffusion is low.  Here, we first derive this effect for a general class of Langevin systems using a response-field path integral framework that makes clear the relationship between the reduced Fokker-Planck parameters and the absorption and thermalization of drive energy in the fast steady-state.  Some related path-integral system reduction techniques have been studied before (e.g., \cite{feynman1963influence_func,bravi2017sub_network}), but in substantially different contexts. Having established a means of explicitly calculating the parameters of the multiplicative noise stochastic process governing the slow variables, we then proceed to analyse the implications for what we term ``least rattling feedback," in which slow variables dynamically finely-tune themselves to bring about fast variable steady-states that attenuate force fluctuations so as to lower the slow variable effective temperature.

The tendency of slow variables in driven systems to move thermophoretically towards regions of lower effective temperature has been noticed in the past, most commonly in situations where the slow variables find a way to reduce the influx of energy from the drive (as in \cite{magiera2015trapping, corte2008chaikin_spun_colloids}.  As we shall see here, however, a striking alternative can arises if the fast variables are capable of exhibiting regular, integrable dynamics; in such a case, least rattling stability can co-exist with strongly coupling to and absorbing work from the external drive.

%This principle can be seen as a generalization of some of the discussion in \cite{}, where a periodically sheered colloid was observed to self-organize into a particular ``quiescent'' configuration.  MAYBE FIND A HOME FOR THIS COMMENT IN DISCUSSION

%In this context, the contribution of this work is two-fold: first, we present a simpler method of integrating out the fast %variables using the response-field path integral framework, and second, we consider systems where the slow coordinates not %only evolve under a conservative coupling to the fast ones, but also feed back onto the fast dynamics. Understanding this %feedback loop -- where the fast variables create an effective nonequilibrium bath for the slow ones, while the slow %variables affect the steady-state behavior of the fast ones -- is the main focus of the present work. 

In the section \ref{sec:anal_slow} of this article, we will present the derivation of our main analytical result, which establishes a relationship between force fluctuations in fast driven variables and the resulting effective temperature experienced by the slow variables in a driven system.  In section \ref{sec:KR_cart}, we will carry out a numerical analysis of the kicked rotor on a cart -- a time-scale separated, damped, driven dynamical system that is ideally suited for demonstrating the predictive power of the ``least rattling'' framework.  Not only will this analysis draw clear connections to methods of equilibrium statistical physics and show how they generalize in such a nonequilibrium scenario, but it will also underline how ``least rattling'' helps to explain the non-trivial relationship between dissipation rate and local kinetic stability in driven systems. 
%Finally in section \ref{sec:discussion} we will argue for the possible relevance of this result for understanding many-body active matter systems.
%In a closing Discussion, we will consider the outlook for applying the approach developed here to the physics of active matter. 

\section{Analytical slow dynamics} \label{sec:anal_slow}
In this section, we lay out a general formalism for extracting slow dynamics in stochastic systems with strong time-scale separation. We will model ``slow" variables $x_{a}$ and ``fast" variables $y_{i}$ that evolve according to a coupled system of Langevin equations.  Our approach will be to integrate out the fast degrees of freedom and develop an effective theory for the dynamics of the slow variables that is controlled by a small number $\epsilon$ which quantifies the time-scale separation between fast and slow.  As we carry out this integration, we will show that the effects on $x_{a}$ from the fast steady-state of the $y_{i}$ variables at leading order in $\epsilon$ are an average force and, more subtley, a random force and renormalized drag that are calculated from the two-point correlation function of the forces acting between $x_{a}$ and $y_{i}$.  These latter effects are identified as an emergent, position-dependent effective temperature experienced by the slow coordinates.

\subsection{Setup}
While the method we present here is not restricted to this context, it is easiest to illustrate on systems whose dynamics can be given by first order equations, as below. In particular, it works the same way for other types of fast dynamics -- such as inertial, or discrete -- as long as there is a fast relaxation to a steady-state. 
\begin{align}
&\eta \,\dot{x}_a = F_a(x_a,y_i,t) + \sqrt{2\,T\,\eta}\,\xi_a \nonumber\\
&\mu \,\dot{y}_i = f_i(x_a,y_i,t) + \sqrt{2\,T\,\mu}\,\xi_i.
\label{eq:general_sys}
\end{align}
Here the noise $\xi$ is usual Gaussian white noise: $\<\xi_\alpha(t)\>=0$ and $\<`\xi_a`(t) `\xi_b`(s)\> = `\delta_a,b` \delta(t-s) $. Taking the limit $\mu/\eta  \equiv \epsilon \ll 1$ amounts to explicitly separating $x_a$  as slow modes, and $y_i$ as fast ones ($a,b,c$ index the slow configuration space, and $i,j,k$ -- the fast one). The natural physical interpretation of this system as overdamped dynamics in a thermal bath of temperature $ T $, with two different damping coefficients $\mu$ and $\eta$, the noise amplitudes given by Einstein's relation, and with the forces $F_a$ and $f_i$ will be implied from now on for concreteness, but is not at all necessary. With a slight adjustment the system could as well represent underdamped dynamics, such as in the kicked rotor model system we characterize below. 

\subsection{Results}
%-cite influence functional
%-characteristic time v. Lyapunov exp?
The detailed derivation of the effective slow dynamics is relegated to Appendix \ref{app:slow_deriv}. Here we mention only the key steps in the derivation. First, we rescale time $t \rightarrow \mu\, t$, making the slow dynamics obey $\dot{x}_a = \epsilon\,F_a + \sqrt{2\,T\,\epsilon}\,\xi_a$, while the relaxation time of fast variables becomes of $ \Ord{1} $. Second, we express probability of slow trajectories in terms of the Martin-Siggia-Rose path integral (also termed the response-field formalism) \cite{cardy2008neq_SM_turbul}, and third, we do a cumulant expansion controlled by $\epsilon$:
\begin{widetext}
\begin{align}
P[x(t)]
&=\frac{1}{Z_x} \int \D \tilde{x} \<\exp{-\int dt \; \left\{i \tilde{x}_a\(\dot{x}_a - \epsilon\, F_a(x,y,t)\) + \epsilon\, T\,\tilde{x}_a^{\;2} \right\}}\>_{y|x(t)} 
\label{eq:MSR_slow} \nonumber\\
&= \frac{1}{Z_x} \int \D \tilde{x} \;\exp{-\int dt \; \left\{i \,\tilde{x}_a\dot{x}_a + \epsilon\, T\,\tilde{x}_a^{\;2} - i \,\epsilon\,\tilde{x}_a \<F_a\>_y + \frac{\epsilon^2}{2} \tilde{x}_a \tilde{x}_b \<F_a, F_b\>_y + \Ord{\epsilon^3}\right\}}.
\end{align}
\end{widetext}
where $ Z_x $ is the normalization, and $ \tilde{x}(t) $ is the auxiliary ``response'' field. In the last line, we see that the $\Ord{\epsilon^2}$ term in the expansion, like temperature $T$, comes in $\propto \tilde{x}^2$, and thus gives a correction to the noise on the slow dynamics -- this is the effect that we will focus on throughout the rest of this paper. Doing this more carefully (as shown in the Appendix) the resulting slow dynamics, which is our main analytical result, are
\begin{align} \label{eq:eff_slow}
&\quad \gamma_{ab}\,\cdot \,\dot{x}_b = \epsilon \<F_a\>_{y|fix\; x} + \sqrt{2\,\epsilon  D}_{ab}\,\cdot\,\xi_b \\
&\gamma_{ab}(x) = \delta_{a,b} + \epsilon  \int dt' \;(t-t')\<i \, \tilde{y}_i\, \partial_b f_i \at{t'}, F_a \at{t} \>_{y|fix\; x} \nonumber\\
&D_{ab}(x) = T \,\delta_{a,b} + \frac{\epsilon}{2} \int dt' \<F_a \at{t'}, F_b \at{t}\>_{y | fix \; x}. \nonumber
\end{align}
where the matrix square root is defined by $B\equiv \sqrt{D} \;\Leftrightarrow\; B.B^T=D$. Dots denote It\^o products, which will be typical here (see sec.\ref{app:noise_corr}). Note that only the connected components of the expectations appear in the expressions for $ \gamma(x) $ and $ D(x) $ (denoted by commas), and thus are insensitive to any deterministic motion of the fast variables.  Further note that there is also an $\Ord{\epsilon}$ correction of the damping coefficient, which, for a fully conservative (undriven) system, matches the noise correction to preserve Einstein's relation, as it must (see sec.\ref{app:equil}). For non-conservative  forces, however, this will not be the case, and the ratio of the effective noise to damping amplitudes can be used to define an effective temperature tensor $T_{eff}(x_a)\equiv \gamma^{-1}.D.\(\gamma^{-1}\)^T$, which will generally depend on the slow coordinates -- i.e., the noise on slow variables becomes multiplicative. 

\subsection{Least Rattling}

The significance of the above formal result is that to extract the effective slow dynamics we need not know everything about the fast modes, but only the mean and variance of the force fluctuations $F_a$ in the $y_i$ (fast) steady-state at fixed $x_a$ (slow d.o.f.). All other details of the fast dynamics become irrelevant by the same mechanism as for the central limit theorem. The slow dynamics thus follow the simple equation \ref{eq:eff_slow}, which can often be solved analytically. Its qualitative behavior is guided by a competition between the mean drift along the average force $ \<F(x)\> $ and a median drift down the effective temperature gradients $ T_{eff}(x) $. While the former effect is larger by a factor $ 1/\epsilon $, it is a vector quantity, and as such, may be suppressed by averaging in case of high-dimensional disordered fast dynamics. This is in contrast to $ T_{eff} $, which comes in as a positive-definite tensor, making it robust to averaging-out. Without rigorously exploring this trade-off for now, in this work we simply choose focus on the effect of $ T_{eff}(x) $, which guides the slow variables towards regions in their configuration space that yield more orderly, less chaotic, or less ``rattling'' fast dynamics (see sec.\ref{app:regularity_Teff}). We suggest that this effect might result in the self-organization lately studied in many non-equilibrium systems \cite{redner2013phase-sep_colloids, schaller2010flock_fillament}.

%median drift
%generality
%how chaotic - T(x): to App
%phase transitions in fast, RG: here

We now expand on a few of the points mentioned above. First, how general is this method? Its scope is basically inherited from the regime of applicability of the Central Limit Theorem (CLT): our requirement of strong time-scale separation amounts to the condition that fast fluctuations decorrelate faster than dynamical time-scale of slow variables. This way their effect on the slow coordinates adds up as i.i.d. random variables, satisfying the conditions of CLT. Thus any fast fluctuations must either decorrelate quickly (e.g., due to thermal noise or chaos) -- thus contributing to the Gaussian noise amplitude, or not decorrelate at all (as with integrable behavior) -- contributing to the mean force $ \<F\> $. This requirement could notably be broken if some fast fluctuations decay slower than exponentially -- a scenario that leads to effective colored noise and anomalous diffusion, but retains much of the general intuition from eq.\ref{eq:eff_slow}.

This framework is particularly useful in cases where fast dynamics can be in several qualitatively different dynamical phases, controlled by the slow variables. E.g., if a fast variable undergoes a transition from chaotic to integrable behavior as a function of some slow coordinate, then we will typically expect its effect to transition from a noise contribution to an average force contribution respectively -- as we will see in the toy system below. Making this precise and describing the relevant universality classes of these transitions based on their symmetry structure can be done within the broader framework of renormalization group flow.  This could allow extracting the effective slow dynamics, much like it allows finding large scale physics for quantum or statistical fields \cite{kardar2007SM_fields}.

Finally, we mentioned above that while the average force $ \<F\> $ causes the \emph{mean} of the $ x_{a} $-ensemble (slow variables) to drift, the multiplicative It\^o noise given by the effective temperature bath $ T_{eff}(x) $, affects only a drift of the \emph{median} of that same ensemble. This latter effect is realized by virtue of the $ p(x_{a}) $ probability distribution growing increasingly heavy-tailed with time (e.g., log-normal distributions are typical), and so while the mean remains fixed, the median will drift towards the low-noise regions. This means that any finite ensemble of trajectories will also settle in the low-noise region, and the mean will never be realized experimentally. Some aspects of this ergodicity-breaking phenomenon were discussed in \cite{peters2013_PRL_ergBreak}, and a similar problem considered in \cite{schnitzer93-Smoluch_for_chemotaxis}. The key for us is that the least-rattling effect is inherently non-ergodic, and is observed only by monitoring the system over time.

\section{Toy Model}\label{sec:KR_cart}
%\subsection{Model Setup}
%the fast dynamics to This is chosen to be flexible in structure to easily explore the qualitatively distinct regimes of slow dynamics, and is thus of more  pedagogic value rather than inherent importance or applicability. A similar model but of more practical relevance is mentioned at the end. To understand the structure and consequences of an effective temperature landscape for the slow variables of a time-scale separated system, it is best first to illustrate and test them on a simple toy model.
To illustrate the above results, we consider a toy model that is designed to be a simplest possible example capturing all the qualitative features we might expect of more general scale-separated driven systems of interest.  Specifically, we take a kicked rotor on a cart setup shown in fig.\ref{fig:cart_charact}a. The fast kicked rotor (Chirikov standard map) dynamics here are chosen as the simplest system that can realize both the chaotic and integrable behaviors under different parameter regimes. Essentially, the system is a rigid pendulum that experiences no external forces except for periodic kicks of a uniform force field (as though gravity gets turned on in brief bursts), and is given by the first two lines in eq.\ref{eq:KR_cart}. We modelled the system to be immersed in a thermal bath by adding a small damping and noise (see third line in eq.\ref{eq:KR_cart}), whose effects have been studied in \cite{feudel1996map100attr, kraut1999KR_noise}. The point relevant for the following analysis is that when the driving force amplitude (henceforth called ``kicking strength") is large, the rotor dynamics are fully chaotic, but if the kicking strength drops below a critical value ($K\lesssim 5$), periodic orbits appear in the configuration space, and are made globally attractive in the presence of damping, thus quickly making the dynamics integrable (we refer to this phenomenon below as ``dynamical regularization"). Thus, by controlling the effective drive strength, it is possible to switch between chaotic and regular regimes of fast dynamics. 
%its pivot were fastened to a rocket periodically turning on

We then fasten the pivot of the fast kicked rotor on a slow cart that can slide back and forth in a highly viscous medium, perpendicular to the direction of the kick accelerations (i.e., along the symmetry axis of the rotor dynamics -- see fig. \ref{fig:cart_charact}a). The cart is pulled by the tension in the rod, which depends on the fast dynamics, while the global cart position $ x $ can feed back on the fast dynamics by having a kicking field that varies along the cart's track $ K(x) $. This way we have slow variables conservatively coupled to driven fast dynamics, and a feedback loop controlled through the arbitrary form of $ K(x) $ -- providing a flexible testing ground. Overall, we argue that, while vastly simplified, this model captures essential physical features of many multi-particle nonequilibrium systems of potential interest.

\subsection{Model Setup}

The toy model explored here is presented in fig. \ref{fig:cart_charact}a: the kicked rotor is attached to a massless cart moving on a highly-viscous track, which ensures that cart's velocity is much smaller than the rotor's. The exact equations of motion for the system can be derived from a force-balance, and in their dimensionless form become:
\begin{align}
& c\,\dot{x} = -\partial_x U(x) + \sqrt{2\,T\,c}\;\xi_x 
+\underbrace{\sin\theta \,\(v^2 -\ddot{x}\,\sin \theta\)}_{\equiv F_x}\nonumber\\
& \dot{\theta} = v \nonumber\\
& \dot{v} =  - K(x)\, \sin \theta \;\delta(t-n) \nonumber\\
&\hspace{7em}- b\, v + \sqrt{2\,T\,b}\;\xi_v 
 - \ddot{x}\,\cos \theta
\label{eq:KR_cart}
\end{align} 
where all lengths are measured in units of rotor arm length, time in units of kicking period, and the angle $\theta$ is $2\pi$-periodic. Note that for practical reasons (see Appendix \ref{app:KR_cart}), we also assumed that the cart is momentarily pinned down during each kick, so as to remove the term $ \frac{1}{2} K(x)\,\sin 2\theta \;\delta(t-n) $ that should otherwise be included in $ F_x $ due to the direct coupling of the kicks to the cart. For now, we can motivate this by saying that the interesting problem is where the driving force affects the slow dynamics only by means of the fast ones, and not directly, while this chosen implementation can simply be viewed as an additional component of the drive protocol. Additionally, to provide more modelling freedom, we can include an arbitrary potential $ U(x) $ acting directly on the cart to produce a conservative force. Time-scale separation in this model implies that back reaction from cart dynamics on the rotor is small -- i.e., here that $\ddot{x} \ll K$ (by differentiating the last line, we see that indeed $\ddot{x} \sim \Ord{v^3/c} \ll 1$ for $ c \gg 1 $). Thus the leading-order feedback from the slow variables onto fast dynamics comes from $ x $-dependence of $ K $, which we have full control over, making for a convenient toy-model. We also independently assume that $b \ll 1$ so that fast dynamics are close to the ideal kicked rotor and retain its features. 

\subsection{Analytical Evaluation}
For large $K$ (above the dynamical regularization threshold, i.e. $ K \gtrsim 5$), the steady-state of the fast dynamics is fully chaotic, and thus thermal -- i.e., we assume thermalization of the entirety of drive energy among the fast fluctuations, as happens in \cite{cohen2013drive_thermalization} for example. This way, the steady-state distribution is Boltzmann, which is here uniform over $\theta$ and Gaussian over $v$, whose variance we can call $ T_R $ (rotor temperature). The symmetry of this state over $ \theta $ and $ v $ gives $ \<F_x\>_{s.s.}=0 $, making the fluctuations dominant. The only remaining parameter we need to find is then $ T_R $, which is fully constrained by energy balance as follows. In general, to keep an ergodic system at an effective temperature that is higher than that of its bath requires dissipation \cite{horowitz2017diss_neq_distr}:
\begin{align}
&\delta Q = \int dt\, v \circ \(b\,v - \sqrt{2\,T_0\,b}\;\xi_v\)
=b \(\<v^2\>-T_0\) \delta t \nonumber\\
&\mathcal{P} \equiv \pd{Q}{t} = b \, \(T_{eff} - T_0\).
\label{eq:power_Teff}
\end{align}
(for 1D systems with mass=1). Moreover, we can find the power exerted by the kicking force to be $\mathcal{P}=K^2/4$ in the chaotic regime, which in the steady-state must balance the dissipated power. This allows us to extract the effective rotor temperature: $T_R \sim T_0 + \frac{K^2}{4\,b}+\Ord{\frac{1}{c}}$ (see sec.\ref{app:KR_ss} for details).

This, however, only gives us information about the fast behavior, while the $x$-noise correction that we want will also depend on the nature of the rotor-cart coupling. This way, we need to evaluate the $x$-force correlations and $\delta T_x =\frac{1}{2 c} \int dt \<F_x(t),F_x(s)\>$, where as above, $F_x = v^2 \,\sin\theta -\ddot{x}\,\sin^2 \theta$ = (centripetal $F_c$) -- (inertia $F_i$) is the force on the cart. The calculation is relatively straightforward and detailed in sec.\ref{app:KR_noise}, where we also show that $\gamma$ damping-coefficient correction is 0 by symmetry of the $ (\theta,v) $ distribution. We find that, while the inertial term can be ignored at leading order, the correlations of the centripetal force give us $\delta T_x = \frac{1}{2c}\int dt \<F_c,F_c\> = K^2/16c$. Note here that this multiplicative noise correction should be interpreted in the It\^o sense, as the $ \<F_c,F_c\> $ correlations decay on a time-scale faster than kick-period (see Appendix,\ref{app:noise_corr}). 

For $ K \lesssim 5 $, on the other hand, the rotor moves periodically in one of the regular attractors. This means that the cart experiences no additional stochasticity other than that from the thermal bath, giving a low $ T_{eff}=T_0 $, but as some of these attractors spontaneously break the left-right symmetry of the problem, we get $ \<F_x\>_{s.s.} \neq 0 $. As the motion in most of these attractors is very simple -- $ n \in \mathbb{Z}$ full revolutions of the rotor per kick -- we can estimate this force explicitly: $ \<F_x\>_{ss} = \int_0^1 dt \; v^2(t)\, \sin \theta(t) \sim b\,v_n + v_n^3/2c $, where $ v_n \equiv 2\pi n $, and $ \theta(t) $ and $ v(t) $ were estimated by integrating the equations of motion \ref{eq:KR_cart} at leading order (see sec.\ref{app:KR_ordered}).

Compiling the resulting predictions for the cart motion, we get:
\begin{align} \label{eq:cart_result}
c\, \dot{x} =& -\partial_x U(x) + \<F\> + \sqrt{2\,c\,T_{eff}}\cdot \xi\\ 
 &\<F\> = \begin{cases}
v_n \(b + v_n^2/2c\) & K\lesssim 5\\
0 & K\gtrsim 5
\end{cases}\nonumber\\
 &T_{eff} = \begin{cases}
T_0 \quad \\
T_0 + K^2/16c \quad 
\end{cases} \nonumber
\end{align}
with $ v_n\equiv 2\pi n $ and $ n $ some random integer, typically smaller than $ \Ord{\sqrt{T_R}} $ (since the rotor first explores its phase-space thermally before finding one of the regular attractors). Another quantity we can easily estimate for the two phases is the energy dissipation rate:
\begin{align}
\dot{Q}=\begin{cases}
v_n^2 \(b + v_n^2/2c\)& K\lesssim 5\\
K^2/4 & K\gtrsim 5
\end{cases}
\end{align}
Numerical simulations confirm these predictions in fig.\ref{fig:cart_charact} c, d, and e respectively.

\subsection{Numerical Tests}

\begin{figure} 
	\includegraphics[width=0.4\textwidth]{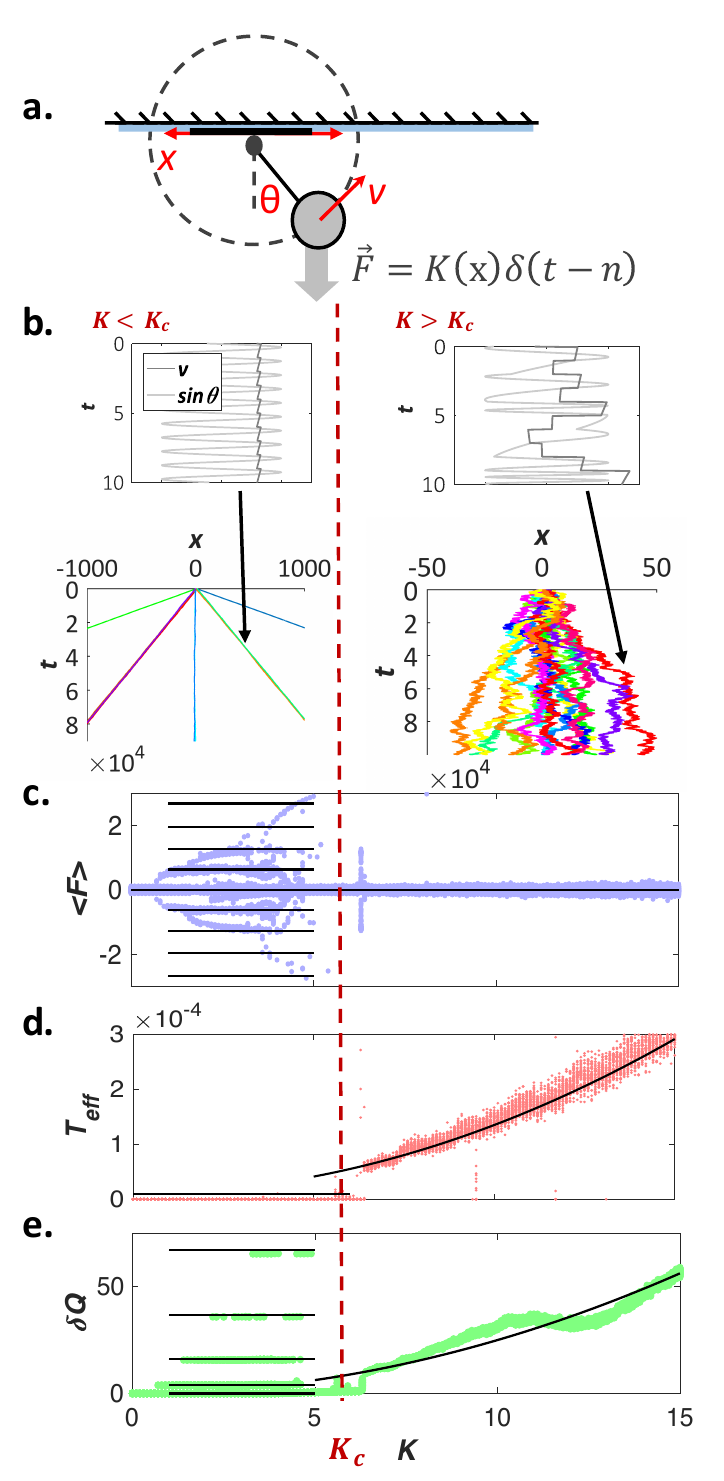}
	\caption{a. schematic of the kicked-rotor-on-a-cart toy model. b. numerical realizations of typical cart trajectories over time for driving force below ($ K=3 $: regular / ordered regime) and above ($ K=8 $: chaotic regime) the ordering transition (at $ K_c \sim 5 $), along with samples of the corresponding fast $ (\theta,v) $ dynamics. c,d,e. average force, fluctuations, and dissipation rates in the cart dynamics, measured from trajectories as in panel (b), for the various values of $ K $, along with analytical predictions (in black) from eq.\ref{eq:cart_result}} \label{fig:cart_charact}
\end{figure}

To verify the above analytical results, we can run numerical simulations of the full system dynamics in eq. \ref{eq:KR_cart}. To begin, we check the cart dynamics for different values of ($ x $-independent) $ K $ (and $ U(x)=0 $). Fig. \ref{fig:cart_charact}b shows typical cart trajectories for $ K $ in the regular and chaotic regimes. More systematically, plotting the apparent average drift $ \<F\> $ and fluctuations $ T_{eff} $ for multiple realizations at each $ K $, we get plots in c and d of fig. \ref{fig:cart_charact} respectively. We thus see quantitative agreement between the prominent features of these plots and the results of eq. \ref{eq:cart_result} -- shown here as black lines. Finally fig. \ref{fig:cart_charact}e shows the heat dissipation rate in the different possible steady-states, showing that while lowering $ T_{eff} $ corresponds to decreased dissipation within the chaotic phase, this rule is violated if we enter a regular dynamic attractor.

Note that as the original problem is stated exactly, and our method allows for full analytical treatment of the slow variables, there are no fitting parameters in any of the curves we are comparing against throughout the numerical study. We use $ c=5\times 10^4,\, b=0.1 $ for all simulations, and to emphasize the effects from fluctuations of the fast variables, we take the actual thermal bath to be at a vanishingly low temperature $T_0 \sim 0$, unless otherwise stated.

\begin{figure} 
	\includegraphics[width=0.5\textwidth]{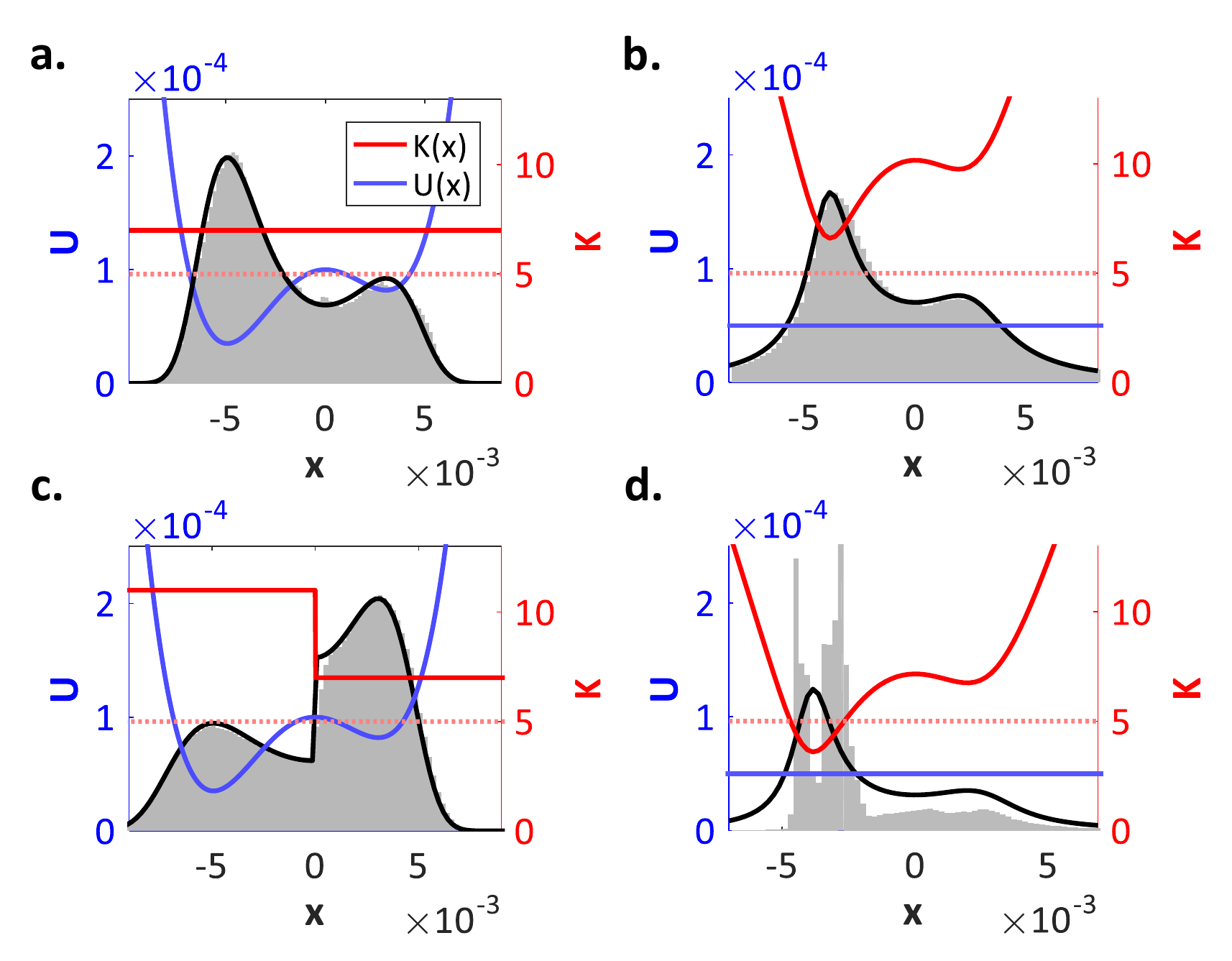}
	\caption{Histograms (grey) of the steady-state cart positions in simulation with: a. constant $K(x)=K$ and in a potential landscape $U(x)$ plotted in blue (black curve gives the expected Boltzmann distribution); b. $U(x)=const$ and $K(x)$ landscape in red (black curve shows $ 1/T_{eff}(x) $ -- solution of the Fokker-Planck equation) c. $U(x)$ as in a. and $ K(x) $ step-function (analytical prediction plotted in black is given by eq.\ref{eq:2-T-wells}). d. $U(x)=const$ and $K(x)$ as in b., but shifted down to dip below the critical $ K_c\sim 5 $ value -- dotted red line (black curve again shows $ 1/T_{eff}(x)$ outside of ordered region) -- this shows that probability gets localized at the two transition points at long times.} \label{fig:distributions}
\end{figure}

While fig. \ref{fig:cart_charact} shows agreement of one- and two-point functions of cart position with our analytical prediction, we have yet to check that the fast dynamics can really be approximated by an effective thermal bath. One convincing way to do this is to introduce a non-trivial potential landscape $U(x)$ acting on the cart's position $ x $, and check the resulting steady-state distribution $p(x)$ against Boltzmann statistics at the predicted temperature $ T_{eff} $. Figure \ref{fig:distributions}a shows the agreement between the histogram produced by this simulation and the curve for the expected Boltzmann distribution.  

To see that the $ T_{eff}(x) $ landscape remains the appropriate description even for non-uniform $ K(x) $, we can calculate the steady-state distribution in a $K(x)$ landscape, now letting $U(x)=0$. The expected distribution for free diffusion in a temperature landscape can easily be found using, e.g., Fokker-Planck equation, and gives $p(x)\propto 1/T(x)$ (note that this arises precisely because our effective slow dynamics have It\^o multiplicative noise -- for Stratonovich it would be $ 1/\sqrt{T(x)} $). This is well confirmed by simulations in fig. \ref{fig:distributions}b, thus showing that at least in the steady-state, probability density does indeed collect in low-temperature regions. 
%(Note that, this tendency is polynomial in $ T_{eff}(x) $, and thus is generally weaker than the tendency to collect in potential wells, which is exponential in $ U(x) $ -- reflecting the )

The last natural test that we mention here is to see how $ T_{eff}(x) $ landscape can counteract the forces of $ U(x) $ -- specifically changing the stability in a double-well potential. This setup is shown in \ref{fig:distributions}c, where the higher-energy potential well is stabilized by having a lower $ T_{eff} $. The numerical result is correctly predicted by the steady-state solution of the Fokker-Planck equation with the expected effective temperatures in each well (labels $L$ and $R$ denote left and right wells respectively), as shown in fig. \ref{fig:distributions}c:
\begin{align}
p(x) = & \frac{1}{Z}\begin{cases}
\frac{1}{T_L}\e{-U(x)/T_L} \qquad &x<0 \quad (L)\\
\frac{1}{T_R}\e{-U(x)/T_R+\Delta} \qquad &x>0 \quad (R)
\end{cases} \label{eq:2-T-wells}\\
& \Delta \equiv U(0)\(\frac{1}{T_R} - \frac{1}{T_L}\) \nonumber
\end{align}
In the limit of a discrete jump process between the two wells (wells with equal internal entropy separated by a high barrier), this exact solution becomes well approximated by that obtained from current-matching with the expected jump rates: $r_\rightarrow = \e{-(U(0)-U_L)/T_L}$ and $r_\leftarrow = \e{-(U(0)-U_R)/T_R}$. The key non-equilibrium feature in these solutions is the dependence of the probabilities in either well on the barrier height $U(0)$ via $\Delta$ -- for higher barriers the temperature difference becomes more important. This example gives the first non-trivial application of thermodynamic intuition from $ T_{eff}(x) $ landscape to solution of our non-equilibrium system. Projecting this concept onto a broader context, we note that this setup is a particular realization in the class of problems of iterative annealing (e.g., used in chaperoned protein folding \cite{todd1996chap_protein_fold}, etc). 

%For the first of these, we consider a double-well potential $U(x)$, with drive amplitude $K$ being different in the two wells, as shown in fig. \ref{fig:distributions}c, along with simulation results. This is similar to such problems as iterative annealing (a known model of chaperoned protein folding, or DNA kinetic proofreading) \todo{examples and references}. The probability distribution  We stress here that in realistic applications, rather than the drive amplitude being different in the two wells, we would expect the fast-variables response to the drive to be more stochastic in one well, as for a misfolded protein configuration, leading to the same effect discussed here. \todo{double well with with one well regularized?}

\subsection{Least rattling}

\begin{figure} 
	\includegraphics[width=0.5\textwidth]{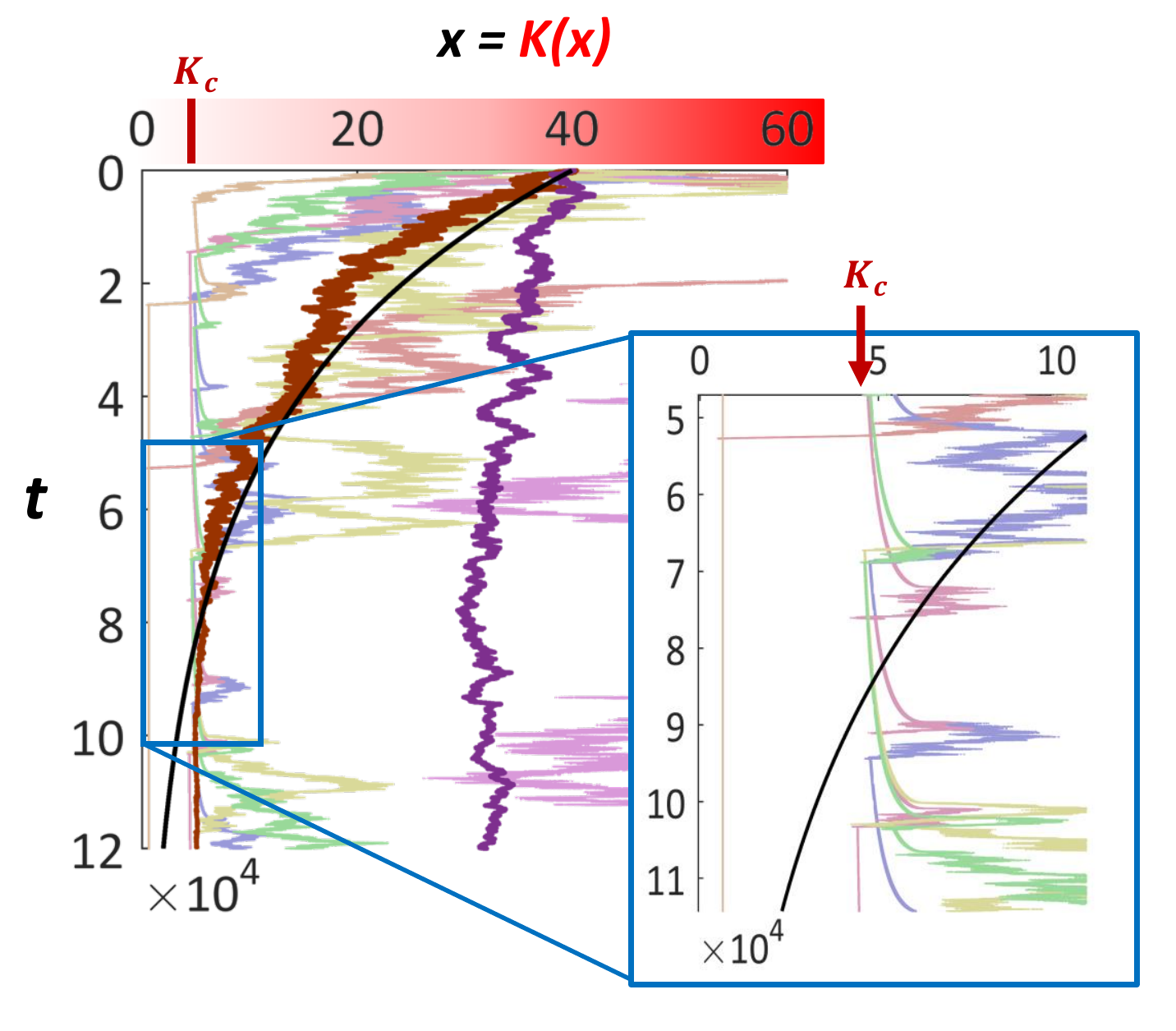}
	\caption{Typical cart trajectories in linear $ K(x) $ landscape ($ U(x)=const $) all starting from one point, along with their mean (purple) and median (brown). Black curve shows the analytical prediction for the median, while mean is expected to be constant at small times. Inset shows the regularization transition at $ K_c \sim 5 $, where effective temperature drops abruptly to 0, and median departs from the smooth decay. $ x $-axis is labelled in units of $ K $} \label{fig:drift}
\end{figure}
Having confirmed the steady-state and thermal properties of the slow behaviors, we next want to look at the predictive power of our formalism for transient behaviors and currents, again in the presence of inhomogeneous fast dynamics. The first example we consider is transient cart motion in linearly varying $K(x)=\kappa\,x$. The simulation results are shown in fig. \ref{fig:drift}. As mentioned above, free diffusion in a temperature gradient results in a median drift to low $T$, as observed here. Explicitly, the slow dynamics in this case $ c\,\dot{x}=\frac{\kappa}{2\sqrt{2}}x \cdot \xi $ can be solved exactly to give $ \ln x(t) = -\(\frac{\kappa}{4c}\)^2\,t + \frac{\kappa}{2\sqrt{2}} \, \mathcal{N}(0,t) $ (with $ \mathcal{N} $ giving the normal distribution with variance $ t $), from which we can read off the mean $ x(t) = x_0 $ and median $ x = x_0\,\exp{-t\,\(\kappa/4c\)^2} $ behaviors. The latter is plotted in black in fig.\ref{fig:drift} and well reproduces the simulation result in brown. Note that for any finite ensemble of trajectories, or for a bounded system, the mean will eventually go to low temperatures as well, but not as cleanly or predictably -- so the constant mean value is not practically realizable at long times. The inset focuses on the crossover into regular dynamics, where we see that the symmetry-broken drift-force $ \<F_x\> $ can either take the cart to the $ K=0 $ absorbing state (as detailed in the further inset), or back out into the chaotic regime. In the latter case, the cart typically diffuses back down to the transition again. The resulting oscillations cause a (transient) accumulation of probability around the critical point, giving a peculiar realization of self-organized criticality. This critical region itself is also interesting as the correlations of the fast variables persist for long times, and can thus break the time-scale separation assumption -- but this will have little effect on the global system behavior. The overall takeaway here is the emergent ``least rattling:'' slow dynamics drift towards regions where fast ones are less stochastic.

To further illustrate the importance of the regularization transition on the slow dynamics, we consider the probability distribution $p(x)$ in the presence of $K(x)$ landscape (and no potential $U=0$), as in fig. \ref{fig:distributions}b., but shifted down such that it dips slightly below the regularization transition at its lowest point -- fig. \ref{fig:distributions}d. The resulting small zero-temperature region in $x$, corresponding to integrable fast dynamics, becomes absorbing, collecting most of the probability density over time (see fig. \ref{fig:distributions}d). Note again that probability accumulates at the critical transition points, giving the two-pronged shape. We stress here the observed sharp localization transition of the steady-state distribution as soon as some regular regime of the fast dynamics becomes accessible -- i.e., the slow variables find the regularized region even if it requires some fine-tuning. (This trade-off between least-rattling and entropic forces can be made quantitative.)

%, which is why in fig. \ref{fig:distributions}d, we plot distributions at two different times, rather than the steady-state. Here we clearly see the accumulation of probability at criticality with the 3-pronged distribution that emerges \todo{better plot, bath noise, does 3-pronged structure persist at long times?}. Finally, we stress that there is a sharp localization transition of the steady-state distribution as soon as some regular regime of the fast dynamics becomes accessible -- i.e., the slow variables find the regularized region even if it requires some fine-tuning. The trade-off between least-rattling and entropic forces can be made quantitative. 
%Note that the three observed sharp peaks within the regular region correspond to three qualitatively different integrable attractors that fast variables can settle into. Finally, we stress that there is a sharp localization transition of the steady-state distribution as soon as some regular regime of the fast dynamics becomes accessible, meaning that the regular regions turn out to be the long-time attractors of the dynamics.

\subsection{Anomalous diffusion}
%Note that both of these examples rely only on non-trivial potential and effective temperature landscapes, and thus would work the same way for a Brownian particle in a potential and heterogeneous temperature bath. However, while for the Brownian particles this situation would be somewhat contrived, we argue that in systems with time-scale separation, such a setup is natural and even typical due to the different possible regimes of coupling to the drive.
%In closing, we wish now to consider a few more intricate examples that could be closer related to potential applications of the formalism to real problems of interest. 

\begin{figure*} 
	\includegraphics[width=1\textwidth]{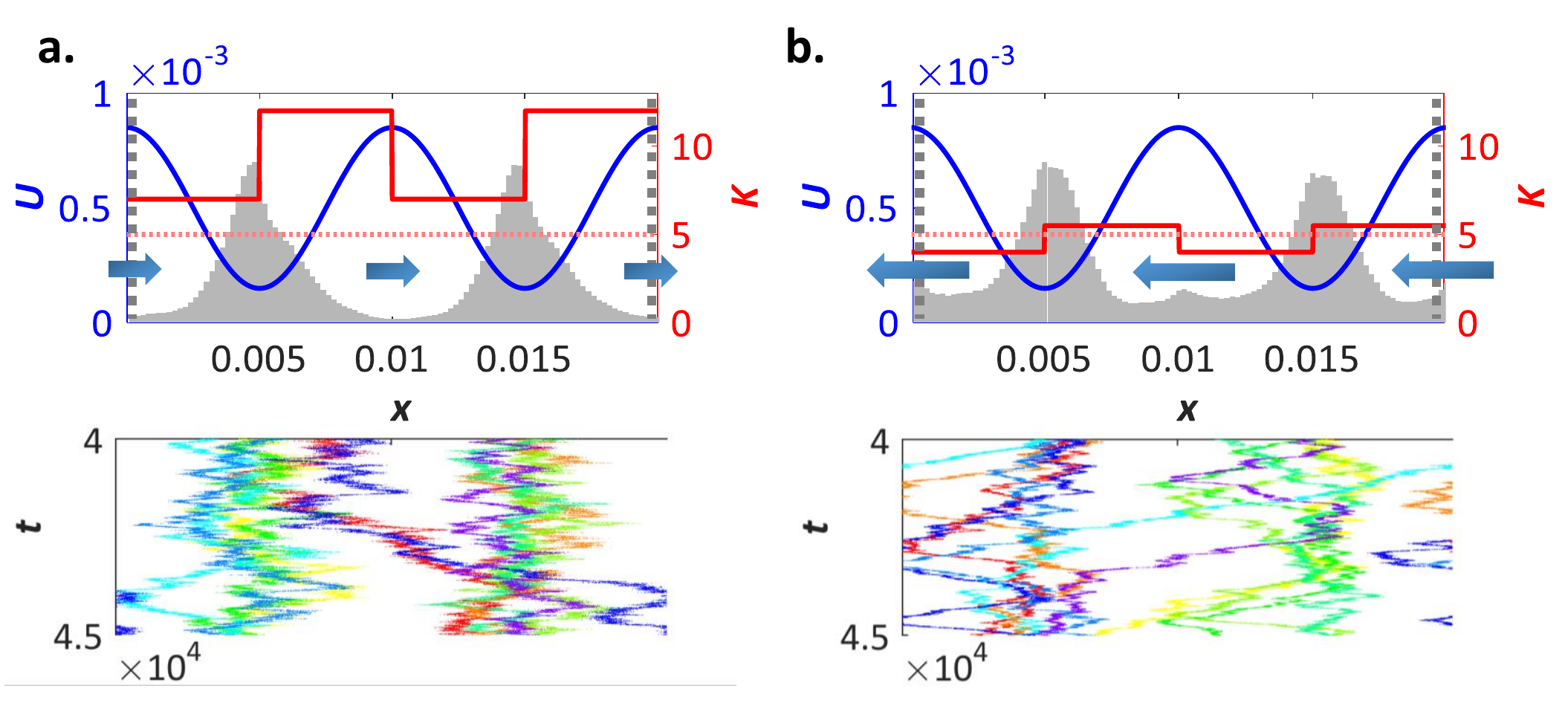}
	\caption{Simulated steady-state cart-position distributions for the shown $ U(x) $ (blue) and $ K(x) $ (red) landscapes ($ x $ is periodic). These result in a pumped steady-state current (block arrows), with the typical barrier-crossing trajectories shown at the bottom. Unlike in all the above simulations, thermal bath temperature $ T_0 = 10^{-4} > 0 $ was taken in these to smooth out the critical behavior. Straddling the critical point with $ K(x) $ in panel b produces a ten-fold larger (and reversed) current, even for smaller absolute variation in $ K(x) $} \label{fig:pump}
\end{figure*}

The last example we present shows that besides an effective temperature landscape, the regular dynamical phase accessible to this model can give rise to apparent anomalous diffusion. To begin, Fig. \ref{fig:pump}a shows an implementation of Buttiker-Landauer ratchet using our model: periodic $ U(x) $ and $ K(x) $ landscapes, with a relative phase-offset of $ \pi/2 $ create a steady-state current being pumped, in this case to the right. Intuitively, this happens because a higher effective temperature in the right half of the potential well makes it easier for the cart to overcome the right potential barrier than the left one. The interesting behavior appears when we shift the $ K(x) $ wave downward to straddle the transition point at $ K \sim 5 $ (fig. \ref{fig:pump}b). In this case the pumped current reverses direction and becomes an order of magnitude \emph{larger} -- even if we reduce the amplitude of the $ K(x) $ variation. To understand this, it helps to look at some typical realizations of barrier-crossing trajectories at the bottom of fig. \ref{fig:pump}. While in panel a. transitions are achieved by stochastic fluctuations that are exponentially suppressed by the Boltzmann factor, in panel b., these are achieved by a directed symmetry-broken drift force $ \<F\> > \partial_x U$, and thus the crossing probability is just the probability of the fast dynamics finding the appropriate regular attractors. These ballistic-like trajectories of the cart in the regular regime can be usefully thought of as anomalous super-diffusion with exponent $ \alpha =2 $. Also, in so far as the barrier crossing becomes easier as we lower $ K(x) $ through the critical value, we can say that the diffusion becomes stronger, thus showing non-monotonicity with $ K $  -- reminiscent of the findings in \cite{spiechowicz2017non-monot_diffusion}. 

\section{Discussion} \label{sec:discussion}
%-connect with Crook's thermal geometry metric\\
%-relation to RG and collective slow d.o.f.
%-connect to mutliple scale analysis, rather than ``removing time'' - splitting time...
The equilibrium partition function that is computed for the Boltzmann distribution is a powerful formal tool for making predictive calculations in thermally fluctuating systems.  Its success stems from two key simplifying assumptions: first, that energy only enters or leaves the system of interest in the form of heat exchanged at a single temperature, and second, that the system and surrounding heat bath uniformly sample joint states of constant energy.  This latter ergodic assumption essentially amounts to eliminating time from the picture, so that energy and probability become interchangeable.

The nonequilibrium scenario is generally less tractable than its equilibrium counterpart both because time has not been eliminated from our description of the system, and also because energy is permitted to enter and leave the system via different couplings to the external environment. Thus, the specific approach to modelling some nonequilibrium systems we have described here seeks to recover some of the desirable advantages of the equilibrium description by exploiting time-scale separation in two ways: first, by only allowing nonequilibrium drives to couple to a fast subset of variables, and second, by ``partially removing'' time from the picture by replacing the fast variables with a timeless thermal bath approximation. This ``conveys'' the entire time-dependence of the problem into the resulting effective slow dynamics. 
% The only remaining time-dependence in the problem is then on the slow scale, where non-ergodic behavior can arise.
%integrating out these fast variables. As in multi-scale analysis, where we imagine partitioning time into a grid, and treating the fast sub-grid-scale processes separately

Adopting such an approach by no means recovers the simplicity of the equilibrium picture, however, it does give rise to a relatively tractable effective description of the dynamics.  As we have seen, slow variables in such a scenario experience not only a mean force landscape from the steady-state of the fast variables, but also are expected to drift in the direction of decreasing fictitious temperature set by the fast variable force fluctuations. Crucially, the latter effect is non-ergodic, thus somehow capturing the breaking of ergodicity typical of driven dynamics in a simple and tractable picture. 

We have established that this effective picture is quantitatively predictive of the diffusive and stationary behavior of distributions for such slow variables in a simple rotor-on-cart toy model. The tendency of such systems to gravitate to values of slow variables that reduce the effective temperature of fast ones suggests a interesting relationship between dissipation and kinetic stability in driven systems.  Although nonequilibrium steady-states are not in general required to be extrema of the average dissipation rate, it is true that the minimum required dissipation to maintain an effective temperature scales with $ T_{eff} $.  Accordingly, there may be a subset of systems where the drift to lower effective temperature is indeed accompanied by a drop in dissipation. However, for cases where dissipation instead goes to maintaining dynamically regular motions, steady-state behavior might be dominated by a highly dissipative, stable attractor of low $T_{eff}$.

%Another useful connection to mention here is that our effective noise tensor $ D_{ab} $ in eq.\ref{eq:eff_slow} is the same as the thermal geometry metric derived in \cite{zulkowski2012geometry,zulkowski2013neq_geom}, which quantified the amount of energy required to externally move the slow variables. In this sense, drift to lower $ T_{eff} $ can be said to be drift to such locations in the slow configuration space where it is easiest for these slow coordinates to adjust if needed. \todo{not so sure about this paragraph... I think we can lose this}

%At the same time, however, we have also used the mixed chaotic dynamics of the kicked rotor to point out that a drop in effective temperature certainly does not bring about a decrease in the rate of dissipation.  takes place in fast degrees of freedom, force fluctuations that contribute to the effective temperature may drop even as dissipation holds steady or increases.  Because 
Moreover, if fast variables can undergo a dynamical ordering transition that is controlled by the slow coordinates, the corresponding drop in effective temperature can be dramatic. As such, this case opens up the intriguing possibility that dynamical ordering in fast variables might serve as a mechanism for the long-term kinetic stability for slow variables.  Moreover, if dynamical ordering only can occur for rare, finely-tuned choices of slow variables,  this stability could appear as a tendency toward self-organized fine-tuning in the slow-variable dynamics. 

Accordingly, we suggest that an interesting future set of applications for the least rattling approach may lie in the active matter setting, where it is frequently the case that coarse-grained macroscopic features of active particle mixtures relax more slowly than the strongly driven microscopic components.  The diversity of self-organized dynamically-ordered collective behaviors exhibited by such systems is well-known \cite{schaller2010flock_fillament}, and it may be useful to characterize these behaviors in terms of their possibly fine-tuned relationships to driven force fluctuations on the microscopic level.  Future work must focus on generalizing our current approach to modelling the dynamics of such coarse-grained variables.

\appendix
\renewcommand{\thesection}{\Alph{section}} 
\section{Derivation of Effective slow dynamics} \label{app:slow_deriv}
%\numberwithin{equation}{section}
\addcontentsline{toc}{section}{Appendices}
%\renewcommand{\theequation}{A.\arabic{equation}}
%\renewcommand{\thesubsection}{\arabic{subsection}} 
%\subsection{Effective slow dynamics} 
Starting with the explicitly time-scale separated dynamics given by the system \ref{eq:general_sys}, the first step is to explicitly bring our small parameter $\epsilon \equiv \mu/\eta \ll 1$ into the equations by rescaling time $t \rightarrow \mu \, t$, which gives:
\begin{align}
&\dot{x}_a = \epsilon \, F_a(x_a,y_i,t) + \sqrt{2\,\epsilon\,T}\;\xi_a \nonumber\\
&\dot{y}_i = f_i(x_a,y_i,t) + \sqrt{2\,T} \;\xi_i
%\label{eq:}
\end{align}
This will allow us to do a systematic expansion in $\epsilon$ below. For later convenience, let us explicitly introduce the time-scales of slow $\tau_S \sim \Ord{1/\epsilon}$ and fast $\tau_F \sim \Ord{1}$ relaxation. Next we want to integrate out the fast degrees of freedom $y_i$, which we can explicitly do by writing down the Martin-Siggia-Rose (MSR) path integral expression for the probability of a particular slow trajectory $x_a(t)$, as given in the first line of eq. \ref{eq:MSR_slow}.
%\begin{align}
%P[x(t)] &=\frac{1}{Z_x} \int \D \tilde{x} \,\<\exp{-\int dt \; \left\{i \tilde{x}_a\(\dot{x}_a - \epsilon\, F_a(x,y,t)\) + \epsilon\, T\,\tilde{x}_a^{\;2} \right\}}\>_{y|x(t)} 
%\label{eq:MSR_slowA}
%\end{align}
For clarity of notation, we have represented the full path-integration over the fast dynamics as the average:
\begin{widetext}
\begin{align}
\<\Op\>_{y|x(t)} \equiv \frac{1}{Z_y[x(t)]} \int \D y \, \D \tilde{y} \; \Op \,\exp{-\int dt \; \left\{i \tilde{y}_i\(\dot{y}_i -  f_i(x(t),y,t)\) +  T\,\tilde{y}_i^{\;2} \right\}} \label{eq:ave_y}
\end{align}
\end{widetext}
Note that so far, this is defined for a specific fixed slow trajectory $x(t)$. With this notation set, we now observe that the only $y$-dependence that the average can act on in eq. \ref{eq:MSR_slow} is that in $F(x,y,t)$. Thus, all the other terms can be taken out of the average, while the remaining small $\epsilon \, F_a$ exponent can be treated with a cumulant expansion:
\begin{widetext}
\begin{align}
&\<\exp{i\,\epsilon\int dt \; \tilde{x}_a \, F_a(x,y,t)}\>_{y|x(t)} = \nonumber\\
&\exp{i\,\epsilon\int dt \; \tilde{x}_a(t) \, \<F_a(x,y,t)\>_{y|x(t)} - \epsilon^2 \int dt \int dt' \;\tilde{x}_a(t)\,\tilde{x}_b(t')\, \<F_a(x,y,t),F_b(x,y,t')\>_{y|x(t)} + \Ord{\epsilon^3}}
\label{eq:cumulant_F}
\end{align}
\end{widetext}
As this is the only part of the path integral in eq \ref{eq:MSR_slow} that carries the coupling to fast dynamics, it will source all the interesting emergent effects (i.e., coupling renormalizations) for the slow modes, and we thus focus on this for most of this Appendix. 

\subsection{Averages over fast dynamics}
%In so far as evaluating this requires the full knowledge of fast evolution, we have not yet simplified the problem.
Before getting into the physical implications of the different terms in the expansion, let's discuss how to go about calculating the averages $\<\Op\>_{y|x(t)}$. Indeed, as defined in eq. \ref{eq:ave_y}, these averages are just short-hand for path integrals over the full fast dynamics in the presence of arbitrarily time-varying slow variables $x_a(t)$, and hence at this point, merely formal, but not very informative, quantities. On the other hand, intuitively we know that at the lowest order in $\epsilon$, these averages should reduce to averages over the $y$-steady states under fixed $x$: $p_{ss}(y|x)$. To derive this result as well as the first correction in $\epsilon$, we need to once again develop a systematic expansion. Besides $\Op$ (see below), the only dependence on the trajectory $x(t)$ in eq \ref{eq:ave_y} comes in through the force $f_i$ (and similarly in the partition function), which by time-scale separation assumption, we know will vary only slightly on the fast time-scale $\tau_F$: $f_i(x(t),y,t)=f_i(x(t_0),y,t) + (t-t_0)\,\dot{x}_a(t_0) \,\partial_a f_i(x(t_0),y,t) +\Ord{\epsilon^2}$. Plugging this expansion into eq. \ref{eq:ave_y} and Taylor expanding both numerator and denominator (normalization $Z_y$) in $\epsilon$, we get:
\begin{widetext}
\begin{align}
\<\Op\>_{y|x(t)} &= \<\Op\>_{y|x(t_0)} + \<\Op, \int dt \; i \tilde{y}_i\, (t-t_0)\,\dot{x}_a(t_0) \,\partial_a f_i(x(t_0),y,t)\>_{y|x(t_0)} + \Ord{\epsilon^2} \nonumber\\
&= \<\Op\>_{y|x(t_0)} + \dot{x}_a(t_0) \int dt \, (t-t_0)\<\Op, \, i \,\tilde{y}_i\, \partial_a f_i \at{t}\>_{y|x(t_0)} + \Ord{\epsilon^2} \label{eq:ave_ySS}
\end{align}
\end{widetext}
%where the second line was derived by extracting all the $y$-independent terms from the average and integrating by parts. 
The averages here $\<\Op\>_{y|x(t_0)}$ are at a fixed $x$, and thus are precisely the averages over $y$ steady-states $p_{ss}(y|x(t_0))$. Note also that at this order, the possible $x$-dependence of $\Op$ is accounted for at the slow time-scale and does not give any additional contributions here. At the next order in $\epsilon$, however, the variations of $\Op$ and $f_i$ on the relaxation time-scale $\tau_F$ begin to interact, giving new contributions. Generally, Feynman diagrams are the only practical way to go to higher orders as the number of correction terms potentially becomes large. We do not employ diagrams here because they are not practical for the general context we are working with -- but they do become very useful in specific examples. 

Applying the result in eq. \ref{eq:ave_ySS} to our cumulant expansion \ref{eq:cumulant_F}, we get 
\begin{widetext}
\begin{align}
%i\,\epsilon\int dt \; \tilde{x}_a(t) \,
\<F_a(x,y,t)\>_{y|x(t)} = \<F_a \at{t}\>_{y|fix\;x(t)} - \dot{x}_b(t) \int dt' \, (t-t')\<\, i \,\tilde{y}_i\, \partial_b f_i \at{t'}\, ,\, F_a \at{t}\>_{y|fix\;x(t)} + \Ord{\epsilon^2}
\label{eq:damping_corr}
\end{align}
\end{widetext}
Remembering the form of eq. \ref{eq:MSR_slow}, we recognize the correction term here as a correction (or renormalization) of the damping coefficient of the original slow dynamics. Crucially, this correction comes in at the same order in $\epsilon$ as the $\Ord{\epsilon^2}$ term in eq. \ref{eq:cumulant_F}, and must thus be kept in our expansion. By the same token, the equivalent correction of the $ \<F_a(t),F_b(t')\>_{y|fix \; x(t)} $ term only comes in at a higher order and is thus ignored at this stage. Nonetheless, it is interesting to note that including higher order corrections would also introduce an inertia-like term into our slow dynamics (mass renormalization, $\propto \ddot{x})$, as discussed in \cite{polkovnikov2016neg_mass}), as well as higher derivative at progressively higher orders.

\subsection{Noise correction}
\label{app:noise_corr}
The key thing to note from the above results is that at $ \Ord{\epsilon^2} $, the only thing we need to compute the slow dynamics are the one- and two-point functions of the various variables in the $y$-steady-states. Thus we need neither the full form of the steady-state distribution of the fast variables, nor the deviations from this steady-state under dynamic $x(t)$. This result should be thought of as (and really is a form of) the central limit theorem. 

Now that we have a sense of what the different terms in the cumulant expansion \ref{eq:cumulant_F} mean mathematically, we can turn to their physical implications. We already mentioned the correction of the $x$-damping coefficient that we get by resolving the $ \<F_a\>_{y|x(t)} $ term in terms of $y$-steady-states. The only other contribution at this same order is the second term in the expansion \ref{eq:cumulant_F}. This will contribute an additional noise term to the resulting slow dynamics, as it will enter the path integral along with $T$, correcting the $\tilde{x}^2$ operator. However, this noise term would only be white if $\<F_a(t),F_b(t')\>_y \sim \delta(t-t')$, which in general need not be the case, hence making the noise correction colored. Intuitively, we see that because of time-scale separation, $y$ fluctuations will decorrelate much faster ($\delta t \sim \Ord{1}$) than the slow time-scale we are sampling by observing $x$ ($\tau_S \sim \Ord{1/\epsilon}$). This makes the short-range correlations of the noise correction unimportant for the slow evolution, allowing us to approximate it by white noise. 

More formally, this situation is precisely identical to having a UV cutoff in a field theory given by, e.g., finite lattice spacing. Similarly, taking the white-noise approximation here corresponds to sending such a cutoff to infinity, which is justified as long as all our observables are confined to energy-scales (or here time-scales) far lower than said cutoff. Explicitly, the approximation we are making (which formally comes from the assumption of RG universality in the fast dynamics):
\begin{widetext}
\begin{align}
\int dt \int dt' \;\tilde{x}_a(t)\,\tilde{x}_b(t')\, \<F_a(x,y,t),F_b(x,y,t')\>_{y|x(t)} \approx 
\int dt \;\tilde{x}_a(t)\,\tilde{x}_b(t) \int dt' \<F_a(x,y,t),F_b(x,y,t')\>_{y|fix \;\hat{x}}
\label{eq:white_approx}
\end{align}
\end{widetext}
% The next question we must address is whether to interpret our $x$-dependent emergent noise term in the It\^o or Stratonovich sense.
One reason why we must be careful in taking this white-noise limit is that the precise limiting procedure will determine whether the correct interpretation of the resulting multiplicative white noise is It\^o or Stratonovich, resulting in observable consequences on the slow time-scale.  This question is related to the choice of $\hat{x}$ where to evaluate the $y$-steady-state in the RHS of eq. \ref{eq:white_approx}, as well as, independently, where to evaluate the explicit dependence $F_a(x)$. To avoid very messy notation, we assume away the latter point by restricting the form of $ F_a(x,y,t) = \tilde{F}_a(x) + F_a(y,t) $, which then makes the above expression \ref{eq:white_approx} depend on $x$ only via the steady-state $p_{ss}(y|\hat{x})$ ($\tilde{F}$ drops out altogether as it only contributes to the disconnected cumulant): define $ \delta T(\hat{x}) \equiv  \frac{\epsilon}{2} \int dt' \<F_a(y,t), F_b(y,t')\>_{y | fix \; \hat{x}}$. Again, this restriction is not necessary and is taken here for convenience.

Thus, we see that if we discretely change $ \hat{x} $, two separate time-scales (both fast, $ \sim \Ord{1} $) control the relaxation of $ \delta T(\hat{x})$: $ \tau_F $, on which $p_{ss}(y|\hat{x})$ globally relaxes to its new form (i.e., relaxation time of one-point functions), and $ \tau_{F2} $, on which the two-point function $\<F_a(y,t), F_b(y,t')\>_{y | fix \; \hat{x}}$ decays. If $ \tau_F \ll \tau_{F2} $, then we have the usual result that the white-noise limit of multiplicative colored noise should be interpreted as Stratonovich (see Ch.6.5 in \cite{gardiner1985handbook}). On the other hand, for $ \tau_F \gg \tau_{F2} $ we see that $ p_{ss}(y|\hat{x}) $ remains essentially fixed while the noise correlations decay, and so the noise amplitude must be evaluated according to the value of $ x $ at the beginning of the $ \tau_{F2} $ interval, i.e., in non-anticipating It\^o convention. Note that while both limits are possible, the latter is typical, especially for many-body systems, since the relaxation of $ p_{ss}(y|\hat{x}) $ proceed via relaxations of two-point functions throughout the system. %is a global consequence of local relaxations of 
Finally, note that the same It\^o / Stratonovich ambiguity occurs in the expression for the damping correction \ref{eq:damping_corr} and is resolved in exactly the same way as here.

\subsection{Compiling results}
Finally, we are in a position to put everything together. We use our final expressions for damping \ref{eq:damping_corr} ($ \delta \gamma_{ab}(x) \equiv \int dt' \, \<\, i \,\tilde{y}_i\, \partial_b f_i \at{t'}\, ,\, F_a \at{t}\>_{y|fix\;x(t)} $), and noise correction \ref{eq:white_approx}
($ \delta T(\hat{x}) \equiv  \frac{\epsilon}{2} \int dt' \<F_a(y,t), F_b(y,t')\>_{y | fix \; \hat{x}} $) in the cumulant expansion \ref{eq:cumulant_F}, and plug that into the full  expression for the probability distribution over slow paths \ref{eq:MSR_slow} to get:

\begin{widetext}
\begin{align*}
P[x(t)] &= \frac{1}{Z_x} \int \D \tilde{x} \;\exp{-\int dt \; \left\{i \,\tilde{x}_a\dot{x}_a + \epsilon\, T\,\tilde{x}_a^{\;2} - i \,\epsilon\,\tilde{x}_a \(\<F_a\>_y - \delta \gamma_{ab}\star\dot{x}_b\) + \tilde{x}_a  \,\delta T_{ab} \,\tilde{x}_b + \Ord{\epsilon^3}\right\}}\\
&= \frac{1}{Z_x} \int \D \tilde{x} \;\exp{-\int dt \; \left\{
	i \,\tilde{x}_a \(\gamma_{ab}(x,t)\star \dot{x}_b - \epsilon\,\<F_a\>_y \) + \epsilon\,\tilde{x}_a \star D_{ab}(x,t)\star\tilde{x}_b + \Ord{\epsilon^3}
\right\}}.\\
&where \qquad 
\begin{aligned}
&\gamma_{ab}(x,t) \equiv \delta_{a,b} + \epsilon  \int dt' \;(t-t')\<i \, \tilde{y}_i\, \partial_b f_i \at{t'}, F_a \at{t} \>_{y|fix\; x} \\
&D_{ab}(x,t) \equiv T \,\delta_{a,b} + \frac{\epsilon}{2} \int dt' \<F_a \at{t'}, F_b \at{t}\>_{y | fix \; x}
\end{aligned}
\end{align*}
\end{widetext}
%where $ \gamma_{ab} $ and $ D_{ab} $ are given in eq.\ref{eq:eff_slow}
The resulting path integral can then be used to extract the corresponding Langevin equation for the slow dynamics:
\begin{align*}
\gamma_{ab}(x)\,\star \,\dot{x}_b &= \epsilon \<F_a\>_{y|fix\; x} + \sqrt{2\,\epsilon  D(x)}_{ab}\,\star\,\xi_b \qquad (It\hat{o})\\
&\star \equiv \begin{cases}
\cdot \quad [It\hat{o}] \qquad & \tau_{F2} \ll \tau_F\\
\circ \quad [Strat] & \tau_{F2} \gg \tau_F 
\end{cases}
\end{align*}
where $\xi_a$ is the usual white noise: $\<\xi_a(t),\xi_b(t')\>=\delta_{ab}\,\delta(t-t')$, and the square root of the matrix $D_{ab}$ is defined by the condition $\sqrt{D}\, .\,\sqrt{D}^T = D$. Finally the notation $\star$ is used to denote the It\^o or Stratonovich dot according to the conditions described in the last section: $\tau_F$ and $\tau_{F2}$ are the decay time-scales for the one- and two-point functions of the fast dynamics respectively. This is then the main analytical result of our work, shown in eq. \ref{eq:eff_slow} for the more common It\^o case.

\subsection{Equilibrium: sanity check} \label{app:equil}
Now that we have the effective slow dynamics for general stochastic systems with time-scale separation, we want to check that in the equilibrium case, we recover the expected fluctuation-dissipation relation: $ D_{ab}(x) = T\,\gamma_{ab}(x)$. Equilibrium in our original system will corresponds to lack of any driving forces: thus all the forces must come from gradients of a single potential landscape $U(x_a,y_i)$: $F_a = -\partial_a U$ and $f_i = -\partial_i U$. Focusing on the expression for $\gamma_{ab}$ above we note that in this case $\partial_b f_i =-\partial_b \partial_i U= \partial_i F_b$
\begin{align*}
\gamma_{ab}(x,t) = \delta_{a,b} + \epsilon  \int dt' \; (t-t')\<i \, \tilde{y}_i\, \partial_i F_b \at{t'}, F_a \at{t} \>_{y|fix\; x} 
\end{align*}
We then note that the response field for the force $F_b$ is given by $\tilde{F}_b = \tilde{y}_i\, \partial_i F_b$ when $x$ is fixed, as it is here. Finally in MSR we know that $ \<i\,\tilde{F}_b,F_a\> $ gives the linear response function for $F$, and so by fluctuation-dissipation theorem $ \<i\,\tilde{F}_b\at{t'},F_a\at{t}\> = \partial_{t'} \<F_b\at{t'},F_a\at{t}\>/T$ for $t'<t$ (zero otherwise). Using this in the above expression and integrating by parts:
%$ \<F_b\at{t'},F_a\at{t}\> = T \int^{t'} d\tau \<i\,\tilde{F}_b\at{\tau},F_a\at{t}\> $ for $t'<t$:
\begin{align*}
\gamma_{ab}(x,t) &= \delta_{a,b} + \frac{\epsilon}{2\, T}  \int dt' \<F_b \at{t'}, F_a \at{t} \>_{y|fix\; x} \nonumber\\
&= D_{ab}/T 
\end{align*}
(The factor of two dividing the integral comes from the fact that while the correlator is time-symmetric, the response function is causal.) We thus recover the desired result. 

\subsection{Fast dynamics and $ T_{eff}(x) $} \label{app:regularity_Teff}
The last question we must address is why does the effective temperature $ T_{eff}(x) $ found above, in general correlate with how chaotic the fast variables are? In the case where our fast dynamics undergo a phase transition, we clearly see that under integrable dynamics (zero Lyapunov exponents), the connected correlator $ \<F_a(y,t) F_b(y,t')\> - \<F_a(y,t)\> \< F_b(y,t')\> = 0$ vanishes (or is proportional to the small thermal bath temperature). By the same token, in the chaotic phase (Lyapunov exponents comparable to inverse characteristic time), the averages $ \<F_a(y,t)\> $ are insensitive to the amplitude of the chaotic fluctuations (by symmetry), and thus we get a high $ T_{eff} $. This is the case in the toy model we studied -- as illustrated in fig.\ref{fig:corr_decay}.

The issue is more subtle, however, when we are not explicitly considering a phase transition in the fast behavior. For example, consider a system that can have chaotic behavior, as well as regular self-oscillations, but with slow random phase-drift. This way, the steady-state probability in both cases is distributed throughout the accessible configuration space, with the only distinguishing feature being the correlation decay time $ \tau_{F2} $ -- scaling inversely with the Lyapunov exponents $\lambda_{Lyap}$. It turns on that in this case also, $ T_{eff} \propto 1/\tau_{F2} \propto \lambda_{Lyap}$ is higher for more chaotic systems -- as long as $ \tau_{F2} > \tau_{char} $ -- characteristic return time of fast dynamics.

We can motivate this claim by first realizing that if the fast steady-state is confined to a finite configuration-space region, then it must have some cyclicity with a finite characteristic return-time $ \tau_{char} $. This means that the correlator $ \<F_a(y,t), F_b(y,t')\>$, besides exponentially decaying, will also fluctuate (not necessarily periodically), with persistence time $\leq \tau_{char}$ (depending on the details of fast-slow coupling). Thus, we can Fourier transform the force correlator as: $ \<F_a(y,t), F_b(y,t')\> \sim \e{-t/\tau_{F2}}\, \int_{\omega_0}^{\infty} d\omega \;f(\omega) \cos(t\,\omega) $, where the infra-red cutoff $ \omega_0 = 2 \pi / \tau_{char} $ is given by the fact that fast dynamics have no time-scales longer than $ \tau_{char} $. Integrating this correlator to recover the effective temperature, we see that $ T_{eff}(x) \propto  \int_{\omega_0}^{\infty} d\omega \;f(\omega)\, \omega^2/\tau_{F2}$ as long as $ \tau_{F2} > \tau_{char} $, as stated above. Of course, all this assumes that the amplitude of the force fluctuations stays roughly the same as their correlation time changes -- but the systems we are interested in are those that exhibit a qualitative change in their Lyapunov exponents, thus making this the dominant effect.

\section{Kicker Rotor on a Cart} \label{app:KR_cart}
In this appendix we present the analytical calculations required to make the predictions for the Kicked Rotor on a cart toy model described in the main text. For convenience, we reproduce the dimensionless equations of motion here (this time including the direct effect of kicks on the cart in $ F_x $):
%\begin{widetext}
\begin{align}
& \dot{\theta} = \underbrace{v}_{\equiv f_\theta} \nonumber\\
& \dot{v} = \underbrace{ - K(x)\, \sin \theta \;\delta(t-n)  
	\color{blue} - \ddot{x}\,\cos \theta
	\color{red}- b\, v  
}_{\equiv f_v} \color{red}+ \sqrt{2\,T\,b}\;\xi_v \nonumber\\
&\color{blue} c\,\dot{x} =  \underbrace{\sin\theta \,\(v^2 + K(x)\,\cos \theta \;\delta(t-n) -\ddot{x}\,\sin \theta\)}_{\equiv F_x} \nonumber\\ 
&\hspace{4em}\color{blue} -\partial_x U(x) + \sqrt{2\,T\,c}\;\xi_x
\label{eq:KR_cart_appx}
\end{align} 
%\end{widetext}
The units were chosen such that rotor arm length, mass, and kicking period are all =1. The part in black gives simple kicked rotor dynamics, red part weakly (for $ b \ll 1 $) couples it to a thermal bath at temperature $ T $, and blue part gives the coupling and dynamics of the cart. We assume throughout that the bath temperature here is very low $ T \sim 0 $ to highlight the effect of the chaotic fluctuations of kicked rotor. Strong time-scale separation, which here is achieved by assuming $ c \gg 1 $, implies that terms $ \propto \ddot{x} $ will be small. To see when precisely we can be justified in dropping these, we estimate their magnitude for the two regimes: the forced regime (``during the kick'') and the free rotation. For the forced regime, since $ \delta(t) $ is a distribution, we can only talk about the integrals
\begin{align*}
\delta v_K = &\lim_{\eta\rightarrow 0}\int_{-\eta}^{+\eta} dt \; \dot{v} 
= -K\,\sin\theta - \cos\theta\,\[\dot{x}\]_{-\eta}^{+\eta}
%\begin{aligned}
%= -K\,\sin\theta - \cos\theta\;\frac{\sin\theta}{c}\[\(v+\delta v_K\)^2-v^2\]\\
\\ 
=&-K\,\sin\theta - \frac{K}{c} \sin^2\theta \,\cos\theta\[K\sin\theta-2\,v\]\\
%\end{aligned}\\
\delta x_K =  &\lim_{\eta\rightarrow 0}\int_{-\eta}^{+\eta} dt \; \dot{x} \\
=&K\,\sin\theta\,\cos\theta - \frac{K}{c}\sin^4\theta \[K\sin\theta-2\,v\]
\end{align*}
For the free rotation, we can simply differentiate the unforced part of the last line in eq. \ref{eq:KR_cart_appx} with respect to time, which gives $ c \,\ddot{x} = v^3 \cos \theta $ to leading order. %+ 2\,v\,\sin\theta\,\(-K\,\sin\theta\,\delta(t-n)\) - K\,v\,\sin\theta\,\delta(t-n)$, since $ \dot{v} $
Thus, to ignore the $ \ddot{x} $ terms, we need $ K\,v/c \ll K $ for the driven regime, and $ v^3/c \ll b\,v $ for free rotation. While the former condition is easy to satisfy for a large $ c $, the latter one competes with our additional assumption that $ b \ll 1 $ and can be difficult to satisfy numerically, especially as velocities $ v $ can sometimes get very large -- thus we will keep the $ \ddot{x}\,\sin \theta $ term as an additional perturbative correction to the free dynamics. % (see below), we must be careful when dropping them.

\subsection{Chaotic Kicked Rotor steady-state}\label{app:KR_ss}
To proceed in evaluating the different terms in the expression for the effective slow dynamics (eq.\ref{eq:eff_slow} in main text), we need to find the steady-state distribution over $ (\theta, v) $ for a fixed cart position $ x $. As mentioned in the main text, for strong driving $ K \gtrsim 5 $, the kicked rotor dynamics are fully chaotic, and thus the steady-state thermalizes all input energy. This immediately implies that the probability distribution is of form $ p_{ss}(\theta,v\,|x) \propto \exp{-\frac{v^2}{2\,T_R(x)}} $ -- uniform over $ \theta $ and Gaussian over $ v $, parametrised by a single number $ T_R(x) $ -- effective rotor temperature. The symmetries of this distribution guarantee that $ \<F_x\>_{ss}=0 $.

To find this temperature, we can use the argument from eq. \ref{eq:power_Teff} of the main text, which tells us that this steady-state will have a dissipation rate:
\begin{align*}
&\delta Q = \int dt\, v \circ \(b\,v - \sqrt{2\,T_0\,b}\;\xi\)\\
&\quad =b \(\<v^2\> - \sqrt{2\,T_0\,b}\; \<\frac{v_{t+\delta t} + v_t}{2}\, \xi_t\>\) \delta t \nonumber\\
\end{align*}
where for underdamped, forced Langevin dynamics, we have in general: \\$ v_{t+\delta t}=v_{t}\, +\, \delta t \(-b\,v_t + F(x,v,t) + \sqrt{2\,T_0\,b}\;\xi\) $. Thus, while $ v_t $ is completely uncorrelated with $ \xi_t $: $ \<v_t\,\xi_t\>=0 $, $ v_{t+\delta t} $ is correlated only via the thermal noise term: $ \<v_{t+\delta t}\,\xi_t\>= \sqrt{2\,T_0\,b} $, and is independent of any interaction or driving forces $ F(x,v,t) $. This gives, for mass=1: 
\begin{align}
&\mathcal{P} \equiv \pd{Q}{t} = b \, \(T_{eff} - T_0\).
\label{eq:power_Teff_app}
\end{align}
(note that if $ v $ were a vector in $ d $-dimensions, we would multiply this expression by $ d $).
With this, and neglecting the bath temperature $ T_0 $, we can balance the work flow in and heat flow out per kick, to get: 
\begin{align*}
0 = &\delta W - \delta Q \\
=& \<\lim_{\eta \rightarrow 0}\int_{-\eta}^{+\eta} dt \; v \circ \(-K\, \sin\theta\)\>_{ss}- b\,\<v^2\>_{ss}\\
= &\<\(v_{pre}-\frac{K}{2}\, \sin\theta\) \(-K\, \sin\theta\)\>_{ss} - b\,T_R \\
&\Rightarrow \quad T_R = \frac{K^2}{4\,b} %+\Ord{\frac{1}{c}}
\end{align*}
where $ v_{pre} $ is the pre-kick velocity, which is uncorrelated with $ \theta $. Since this gives the variance of typical rotor velocities, we can use it to simplify the time-scale separation condition derived above $ v^3/c \ll b\,v $ down to $ K/c \ll b^2 $ -- which is quite difficult to satisfy in addition to $ b\ll 1 $. Thus, while this result is correct, it turns out that $ \Ord{1/c} $ correction coming from the cart coupling term $ \ddot{x} \,\cos\theta $ is very important here. Note that eq. \ref{eq:eff_slow} in main text tells us to include $ x $-dependence of the steady-state in correcting the effective damping coefficient $ \gamma $, however as this term depends only on $ \ddot{x} $ and not $ x $ itself, the steady-state does not gain any $ x $-dependence from it, but rather an $ x $-uniform correction which must be included directly. In this case, as the dynamics are still chaotic and distribution thermal, while the work extracted from the drive $ \delta W $ is not affected (since any work done on the overdamped cart is immediately dissipated and so can be ignored), so the coupling to cart simply adds another channel for heat dissipation:
\begin{align*}
\delta Q = &b \<v^2\>_{ss} + \<\int_{0}^{1} dt \; c\, \dot{x}^2\>_{ss} \\
=&b\, T_R + \frac{1}{c} \int_{0}^{1} dt \; \< F_x\at{t}F_x\at{t}\>_{ss}\\
=&b\, T_R + \frac{1}{c} \Big[\int_{0}^{1} dt \; \< \(v^2\,\sin\theta\)^2 \>_{ss} \\
&\qquad + \< 2\, K\,v^2\,\sin^2\theta\,\cos\theta\at{t=0}\>_{ss}\Big]\\
=&b\, T_R + \frac{1}{c} \[\frac{3}{2}\,T_R^{\;2} + 0\]\\
&\Rightarrow \quad T_R = \frac{b\,c}{3}\(\sqrt{1 + \frac{3\,K^2}{2\,b^2\,c}}-1\) < \frac{K^2}{4\,b}
\end{align*}
where we assumed that $ v $ and $ \theta $ are uncorrelated during most of the time $ 0<t<1 $, as justified below. This correction significantly lowers the typical velocities and well reproduced in the results of the simulations for large, but practical values of $ c $ (e.g., for $ b= 10^{-2},\, c= 10^4,\, K= 10 \Rightarrow T_R\approx 375 < K^2/4b=2500$, or for $ b= 10^{-1},\, c= 5\times 10^4,\, K= 10 \Rightarrow T_R\approx 230 < K^2/4b=250$).
%will be of $ \Ord{K/\sqrt{b}} $, which can get quite large (e.g., $ b=0.01,\, K=10 \Rightarrow v\sim 100 $), thus straining the constraint .
%There is a practical issue with this result, however, coming from the term $ \ddot{x} \,\cos\theta $ as mentioned above (see eq. \ref{eq:KR_cart_appx}). While for sufficient separation $ 1/c \ll b \ll 1 $ this result is correct, such separation is difficult to realize numerically (especially as we want to study effects of $ \Ord{1/c} $). Thus, we cannot neglect 
%: the separate requirement that $b \ll 1$ competes with the time-scale separation condition that $ c \gg 1 $.  requires us to keep the first-order correction in $1/c$.

\subsection{Cart Damping and Noise correction} \label{app:KR_noise}

\begin{figure*}
	\includegraphics[width=1\textwidth]{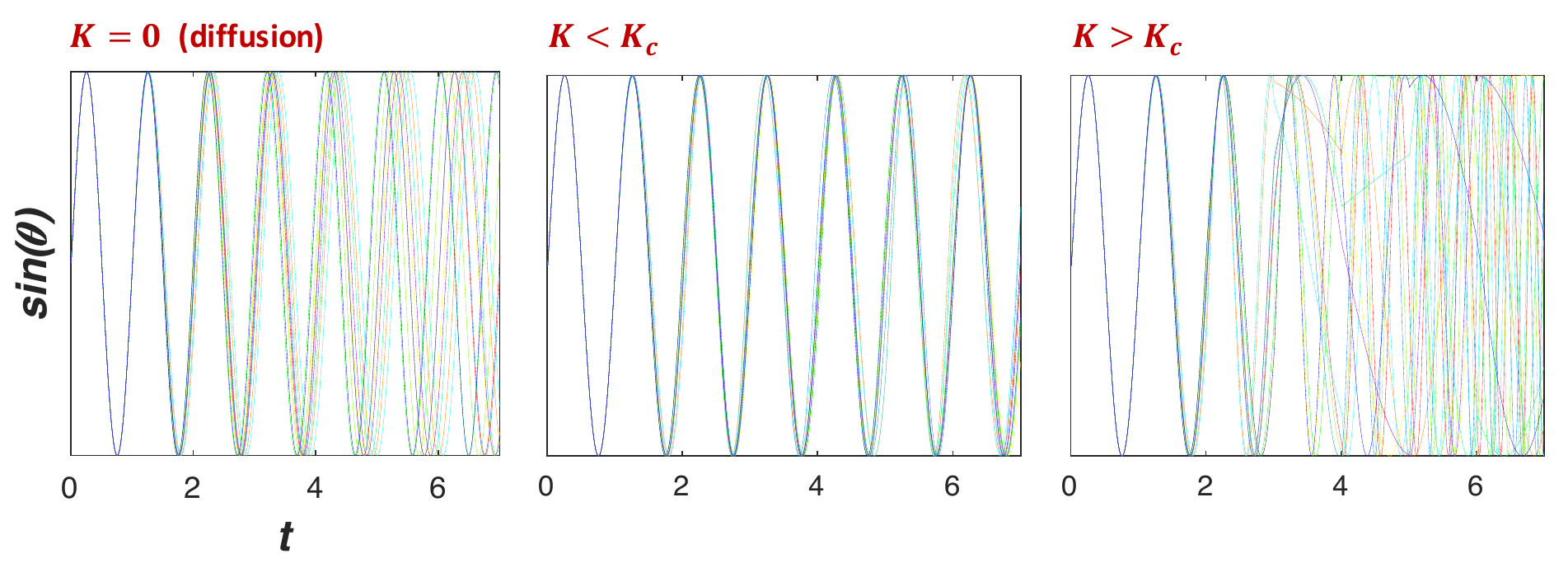}
	\caption{Different realizations of kicked-rotor trajectories starting from same initial conditions in the regular and chaotic driven phases, as well as the freely-diffusing undriven case (for reference). While the steady-state in the chaotic phase looks like large random fluctuations about a constant mean (thus giving a high $ T_{eff} $), in the regular phase, the drive serves to synchronize all trajectories, thus yielding small fluctuations about an oscillating mean in the steady-state (and so a low $ T_{eff} $).}
		%decay of the $ \<F_x \at{t}, F_x \at{t'}\>_{(\theta,v) | fix \; x} $ correlator over time after a kick at 0 (kick period =1). Colored curves -- different numerical realizations, black curve -- analytical prediction.}
	\label{fig:corr_decay}
\end{figure*}

With the above understanding of the steady-state, we now proceed to compute the two-time correlations functions needed to get the corrections on the slow dynamics given by eq.\ref{eq:eff_slow} of the main text. We begin by noting their general structure here: each kick introduces correlations between $ \theta $ and $ v $, after which, while $ v $ remain approximately constant until the next kick (for $ b \ll 1 $), $ \theta $ spins around and correlations decay. The typical decay time-scale in this system can be estimated by looking at the decay:
\begin{align*}
\<\sin\theta(0)\,\sin\theta(t)\>_{ss}
=\<\sin\theta(0)\,\sin\(\theta(0)+v\,t\)\>_{ss}\\
=\frac{1}{2}-\frac{T_R}{4}\,t^2 +\Ord{t^4}
\end{align*}
where we assume $ \theta $ and $ v $ to be uncorrelated over the time-window. This gives decay $ \tau \sim 1/\sqrt{T_R} $, which for typical values of parameters (e.g., for $ b= 10^{-1},\, c= 5\times 10^4,\, K= 10$) could be around 1/20. The key here is that in most cases the decay time is much shorter than 1 (the kicking period). Figure \ref{fig:corr_decay} shows numerical results for $ \<F_x(t), F_x(t')\> $ correlation in the chaotic phase to give a sense of how these quantities typically look for the given system. Since the $ \theta - v $ correlations are only generated by kicks, this implies that most of the time they are uncorrelated -- as we have assumed a few times above. Moreover, this means that our noise and damping corrections should always be interpreted as It\^o for this system, as discussed in the Appendix \ref{app:noise_corr} of the main text.

Using this result we can immediately see that the damping correction vanishes (see eq.\ref{eq:KR_cart_appx} for definitions of $ f_\theta, f_v $):
\begin{align*}
&\delta \gamma =  \frac{1}{c} \int dt' \;(t-t')\<i \, \(\tilde{\theta}\, \partial_x f_\theta + \tilde{v}\, \partial_x f_v\) \at{t'}, F_x \at{t} \>_{(\theta,v)|fix\; x}\\
&=\frac{1}{c} \int dt' \;(t-t')\<-i \, \tilde{v}\, K'(x)\, \sin \theta \;\sum_n \delta(t'-n) \at{t'}, F_x \at{t} \>_{(\theta,v)|fix\; x} \\
&=\frac{1}{c} \;\sum_n (t-n)\<-i \, \tilde{v}\, K'(x)\, \sin \theta \at{n}, F_x \at{t} \>_{(\theta,v)|fix\; x} =0
\end{align*}
%since the correlator is only non-zero for $ \abs{t-n}\ll 1 $, but that same region is suppressed by the $ (t-n) $ prefactor. %<- this is wrong I think 
%More carefully, we can check that the symmetries of the thermal steady-state actually make the correlator exactly zero. 
as the correlator vanishes by the symmetries of the thermal steady-state. 

With that, to find the effective temperature experienced by the cart, we need only compute the noise correction, as: $T_{eff} = \frac{1}{2 c}\int dt' \<F_x \at{t'}, F_x \at{t}\>_{(\theta,v) | fix \; x} $. From eq.\ref{eq:KR_cart_appx}, we get $F_x = v^2\,\sin\theta + K(x)\,\sin\theta\,\cos \theta \;\delta(t-n) -\ddot{x}\,\sin^2 \theta $ = (centripetal $F_c$)  + (direct kick coupling $F_k$) -- (inertia $F_i$). Unlike in $ f_v $, where the term $ b\,v $ was comparable magnitude to $ \ddot{x}\,\cos\theta $, here $ F_c $ and $ F_k $ are both $ >\Ord{1} $, and thus the inertia $ F_i $ is distinctly sub-leading and can be dropped. We now proceed to individually compute the $ \<F_c,F_c\> $, $ \<F_k,F_k\> $, $ \<F_c,F_k\>=\<F_k,F_c\> $ contributions. 

For the $ \<F_c,F_c\> $  term, we see that far from the kicks, where $ \theta $ and $ v $ are uncorrelated, we get: 
\begin{align*}
\int dt' \; \<v^{2} \sin\(\theta+t'\,v\),\, v^2 \,\sin \theta\>_{(\theta,v)|fix\; x} = 0
\end{align*}
which can be evaluated analytically in Mathematica. The leading correction to this quantity then comes from the $ \theta - v $ correlations generated by the kicks. To capture these, we write all velocities and angles in terms of their values before the last kick -- at a time when they were guaranteed to be uncorrelated. Thus 
\begin{align*}
&v(t) = \begin{cases}
v_{pre}  \\
v_{pre}-K\sin\theta_{pre}  
\end{cases}
\qquad \\\text{and} \quad 
&\theta(t) =  \begin{cases}
\theta_{pre} + t \, v_{pre} \quad & t<0 \\
\theta_{pre}+t\,(v_{pre}-K\sin\theta_{pre}) \quad & t>0
\end{cases}
\end{align*}
Using these expressions, we can thus evaluate $ \iint dt \, dt' \<F_c \at{t}\;, F_c \at{t'}\> $ piecewise (where a second integral must be included since the time-translation-invariance is now broken). Dropping the subscript \textit{pre}, we have:
\begin{widetext}
\begin{align*}
&\int_{-1/2}^{1/2} dt \int_{-\infty}^{\infty} dt' \<v(t)^2 \, \sin\theta(t),\; v(t')^2 \, \sin\theta(t')\>_{(\theta,v) | fix \; x}\\
&= \begin{cases}
\<\int_{0}^{1/2} dt \int_{0}^{\infty} dt' \(v-K\sin\theta\)^4 \, \sin(\theta+t(v-K\sin\theta)) \, \sin(\theta+t'(v-K\sin\theta))\>_{(\theta,v) | fix \; x}\\
+ 2 \< \int_{-1/2}^0 dt \int_{0}^{\infty} dt' \;v^2 \, \sin(\theta+t\,v) \, \(v-K\sin\theta\)^2 \sin(\theta+t'(v-K\sin\theta))\>_{(\theta,v) | fix \; x}\\
+ \< \int_{-1/2}^0 dt \int_{-\infty}^{0} dt' \;v^4 \, \sin(\theta+t\,v) \,\sin(\theta+t'\,v)\>_{(\theta,v) | fix \; x}
\end{cases} 
= \begin{cases}
K^2/8 +T_R/2\\
-T_R\\
+T_R/2
\end{cases} =K^2/8
\end{align*}
\end{widetext}
where all integrals can be evaluated analytically (in Mathematica) if we take the time integrals first, and then average over the (uncorrelated) $ \theta $ and $ v $. %The analytical expression for the two-time correlator thus obtained is compared to numerical results in fig.\ref{fig:corr_decay}. 
Similarly for the other terms:
\begin{widetext}
\begin{align*}
&\int dt' \<F_k \at{t}\;, F_k \at{t'}\>_{(\theta,v) | fix \; x}
=\<\int_{-\infty}^{\infty} dt' \int_{-1/2}^{1/2} dt\; \sum_{n,m} \frac{K}{2}\sin 2\theta(t)\;\delta(t-n) \; \frac{K}{2}\sin 2\theta(t')\;\delta(t'-m)\>_{(\theta,v) | fix \; x}\\
&=\sum_{m} \frac{K^2}{4}\<\sin 2\theta(0)\; \sin 2\theta(m)\>_{(\theta,v) | fix \; x} = K^2/8
\end{align*}
\begin{align*}
&\int_{-\infty}^{\infty} dt' \<F_k \at{t}\;, F_c \at{t'}\>_{(\theta,v) | fix \; x}
=\int_{0}^{\infty} dt' \<\frac{K}{2} \sin 2\theta\; \(v-K\sin\theta\)^2 \, \sin(\theta+t'(v-K\sin\theta)) \>_{(\theta,v) | fix \; x}\\
&=-K^2/8 = \int_{-\infty}^{\infty} dt' \<F_c \at{t}\;, F_k \at{t'}\>_{(\theta,v) | fix \; x}
\end{align*}
\end{widetext}
Adding up all four terms, we thus get $T_{eff} = \frac{1}{2 c}\int dt' \<F_x \at{t'}, F_x \at{t}\>_{(\theta,v) | fix \; x} =0$. Here we clearly see that the cancellation comes up due to the anti-correlations between the centripetal force and the kick coupling. In this particular case, the cancellation is somewhat accidental, and is a consequence of the simplicity of the system -- the functional form of couplings is quite restricted. In general, we expect such cancellations to be unlikely in higher-dimensional systems. As discussed in the main text, to make the system interesting and get a finite $ T_{eff} $, we can simply eliminate the direct kick-cart coupling $ F_k $ from the dynamics altogether, with the physical interpretation of ``pinning'' down the cart at the instant of the kick. This leaves only $ F_c $, thus giving $ T_{eff} = K^2/16c $, as desired. 

Note also that the rotor temperature $ T_R $ calculated above ends up dropping out and does not affect any of the time-integrated correlators, but only the particulars of their time-dependence as $ \<F_x(t), F_x(t')\>_{(\theta,v) | fix \; x} $. Thus the only really key role it played for us was to show that these correlators decay faster than kicking period.

\subsection{Ordered KR steady-state} \label{app:KR_ordered}
On the other hand for weak driving $ K\lesssim 5$, the kicked rotor undergoes dynamic regularization, and in steady-state is found in one of the integrable attractors in its phase space. Thus, none of the above arguments apply here. Instead, the main regular regions correspond to the rotor completing $ n $ full revolutions per kick, with $ n=...,-2,-1,0,1,2... $. As it does, there are no stochastic fluctuations, other than those from the thermal bath, and as the steady-state lacks the symmetries of the thermal state, $ \<F_x\>_{ss} \ne 0 $ except at $ n=0 $. In fact, depending on the attractor that the rotor falls into, it will exert a persistent force on the cart, causing constant directed drift. We can easily estimate this drift force for the $ n $'th attractor as (here we again assume that $ b\sim\Ord{1/c}\ll 1 $, and let $ v_n \equiv 2\pi n $):
\begin{align*}
&\<F_x\>_{ss} = \int_0^1 dt \; v^2(t)\, \sin \theta(t) = b\,v_n + \frac{v_n^3}{2\,c} + \Ord{b^2}\\
&\text{where} \quad v(t)= 2\pi n - b\,v\,t - \frac{1}{c}\int_0^t d\tau\; v^3 \cos^2(v\,\tau) + \Ord{b^2}\\
&\quad \text{and} \quad \theta(t)=\int_0^t d\tau\;v(\tau)
\end{align*}
where $ v(t) $ dynamics are found by directly integrating eq.\ref{eq:KR_cart_appx} to first order in small parameters. However, as it is impossible to predict which of the attractors will be chosen, we can't a-priori tell the direction or speed that the cart will be moving at -- though the options are restricted to the above small discrete set of possibilities parametrized by $ n $. 
\\\\\\
ACKNOWLEDGEMENTS: We are grateful to Jordan Horwitz, Tal Kachman, Robert Marsland, and David Theurel for provoking discussions and thoughtful comments on this article.
P.V.C. and J.L.E. are supported by the Gordon and Betty Moore Foundation through Grant GBMF4343. J.L.E. further acknowledges the Cabot family for their generous support of Massachusetts Institute of Technology.

%\nocite{*}
\bibliographystyle{unsrt}
\bibliography{/Users/pchvykov/Documents/GoogleDrive/MIT/RESEARCH/_data/14654301E131WECPZ6ZSATT3H7YHJ7G6D2E8/default_files/non-equilibSM.bib}
%/Users/pchvykov/Documents/Git/kr_cart_paper/non-equilibSM.bib

\end{document}